\documentclass[sn-mathphys]{sn-jnl}

\usepackage{t1enc}
\usepackage{times}              
\usepackage{graphicx}           
\usepackage[ansinew]{inputenc} 
\usepackage{amssymb}
\usepackage{esdiff}
\usepackage{float}

\newcommand{\aap}{{A\&A.}}

\newcommand{\prl}{{Phys.~Rev.~Lett.}}

\newcommand{\nphysa}{{Nucl.~Phys.~A}}

\jyear{2025}%

\theoremstyle{thmstyleone}%
\theoremstyle{thmstyletwo}%

\theoremstyle{thmstylethree}%

\begin{document}

\title[Article Title]{\title{Experiments towards a neutron target for measurements in inverse kinematics}
}


\author[1]{\fnm{S.F.} \sur{Dellmann}}
\author[2]{\fnm{C.M.} \sur{Harrington}}
\author[1]{\fnm{O.R.} \sur{Cantrell}}
\author[1]{\fnm{A.L.} \sur{Cooper}}
\author[1]{\fnm{A.} \sur{Couture}}
\author[1]{\fnm{D.V.} \sur{Gorelov}}
\author[1]{\fnm{I.} \sur{Knapov\'a}}
\author[1]{\fnm{S.M.} \sur{Mosby}}
\author*[1,3]{\fnm{R.} \sur{Reifarth}}\email{rreifarth@lanl.gov}
\author[4,5]{\fnm{A.} \sur{Alvarez}}
\author[6]{\fnm{A.} \sur{Aprahamian}}
\author[3]{\fnm{J.} \sur{Butz}}
\author[6,7]{\fnm{I.J.} \sur{Bos}}
\author[2]{\fnm{M.T.} \sur{Febbraro}}
\author[4,8]{\fnm{T.} \sur{Hankins}}
\author[4,5]{\fnm{B.M.} \sur{Harvey}}
\author[3]{\fnm{T.} \sur{Heftrich}}
\author[4,8]{\fnm{M.} \sur{Le}}
\author[2]{\fnm{J.J.} \sur{Manfredi}}
\author[4]{\fnm{A.B.} \sur{McIntosh}}
\author[6]{\fnm{K.V.} \sur{Manukyan}}
\author[6]{\fnm{M.} \sur{Matney}}
\author[4]{\fnm{S.} \sur{Regener}}
\author[6]{\fnm{D.} \sur{Robertson}}
\author[6]{\fnm{A.} \sur{Simon}}
\author[3]{\fnm{D.} \sur{Sokolovic}}
\author[6]{\fnm{E.} \sur{Stech}}
\author[4]{\fnm{G.} \sur{Tabacaru}}
\author[6]{\fnm{W.} \sur{Tan}}
\author[6]{\fnm{M.} \sur{Wiescher}}
\author[4,8]{\fnm{S.} \sur{Yennello}}

\affil[1]{\orgname{Los Alamos National Laboratory}, \orgaddress{\city{Los Alamos}, \state{NM}, \country{USA}}}
\affil[2]{\orgname{Air Force Institute of Technology}, \orgaddress{\city{WPAFB}, \state{OH}, \country{USA}}}
\affil[3]{\orgname{Goethe University Frankfurt}, \orgaddress{\city{Frankfurt}, \country{Germany}}}
\affil[4]{\orgname{Cyclotron Institute, Texas A$\&$M University}, \orgaddress{\city{College Station}, \state{TX}, \country{USA}}}
\affil[5]{\orgname{Physics Department, Texas A$\&$M University}, \orgaddress{\city{College Station}, \state{TX}, \country{USA}}}
\affil[6]{\orgname{Department of Physics and Astronomy, University of Notre Dame}, \orgaddress{\city{Notre Dame}, \state{IN}, \country{USA}}}
\affil[7]{\orgname{Calvin University}, \orgaddress{\city{Grand Rapids}, \state{MI}, \country{USA}}}
\affil[8]{\orgname{Chemistry Department, Texas A$\&$M University}, \orgaddress{\city{College Station}, \state{TX}, \country{USA}}}



\abstract{
Neutron-induced reactions play an important role in fundamental nuclear physics, nuclear astrophysics, and applications. In the case of reactions on rare isotopes, there are limited options for direct experimental measurements. The Neutron Target Demonstrator project at Los Alamos National Laboratory seeks to test the feasibility of moderating spallation neutrons within a 1~m$^3$ graphite cube to create a standing neutron target for neutron-induced reaction measurements in inverse kinematics. This paper presents the results of experimental neutron flux distribution tests using neutron sources (ranging from 1~keV to 50~MeV) created by accelerators at the University of Notre Dame and Texas A\&M University. Measurements were made with both the full graphite cube as well as a ''half cube'' setup in which half of the graphite cube was removed. The measured distributions agree with simulated distributions in the case of the full cube moderator, although there remain discrepancies in certain cases for the half cube moderator. The results shown here will provide useful information for an upcoming experimental campaign to test the neutron target proof-of-principle.

}




\maketitle

\section{Introduction}\label{sec:introduction}

Roberto Gallino pioneered the idea of analyzing the reaction flow around branch points in the $s$ process to significantly constrain the corresponding models \cite{AKW99b}. His inspiring collaboration with experimental nuclear physicists resulted in many experimental campaigns to determine nuclear reaction rates with high impact \cite{WVA01,RKV04}. One of his repeated demands was experimentally constrained neutron-induced reaction cross sections for radioactive isotopes acting as branch points in the $s$ process \cite{KGB11,RAH03}. The ideas presented in this paper address this need for more and accurate nuclear data. If successful, all branch points in the $s$ process \cite{KTG16,sensitivities_online}, all waiting points in the $i$ process \cite{HPW11,GZY13,sensitivities_online} and even some of the important isotopes during the freeze-out phase of the $r$ process \cite{SuE01} can be \textit{directly} investigated in the future. 

Traditional methods to directly determine neutron-induced reaction rates rely exclusively on the idea of a neutron beam interacting with a sample. Depending on the detection mechanism, different techniques are established. The most important ones for neutron energies above 1~keV are \textit{time of flight}, where promptly emitted particles are detected, and \textit{activation}, which relies on the detection of the radioactive product nuclei or delayed emitted particles \cite{RLK14}. 

Both methods reach their limits at modern facilities once the half-life time of the isotope under investigation is less than about 1~year \cite{CoR07}. The most important reason is the emission of the decaying sample atoms interacting with the detectors, which limits the maximum activity of the sample and henceforth the number of sample atoms. 

An idea to push this limit down by orders of magnitude is to invert the kinematics. In other words, a sample of the light reaction partner will be irradiated with a beam of the radioactive ions. This approach has been successfully applied in many cases where the light reaction partner is stable, either in single-pass mode \cite{LLK23} or in combination with an ion storage ring \cite{DGL25}.

Inverse kinematics for neutron-induced reactions poses the particular challenge of a free-neutron target. Simulations have shown that a combination of a reactor or a spallation neutron source with an ion storage ring provides sufficient luminosity to measure neutron induced reactions above 100~keV in the center of mass  \cite{ReL14,RGH17}. In particular the usage of a spallation neutron source is very promising, since they can be rather easily operated at high-energy beam facilities. In contrast to nuclear reactors, spallation driven neutron sources can be shut down quickly and safely.

So far, all estimates rely solely on Monte Carlo simulations. In this paper, we report about the first experimental campaign to confirm the simulation suite used so far.

\section{A spallation-driven neutron target}\label{sec:neutron_target}

\begin{figure}[H]
	\includegraphics[width=0.9\linewidth]{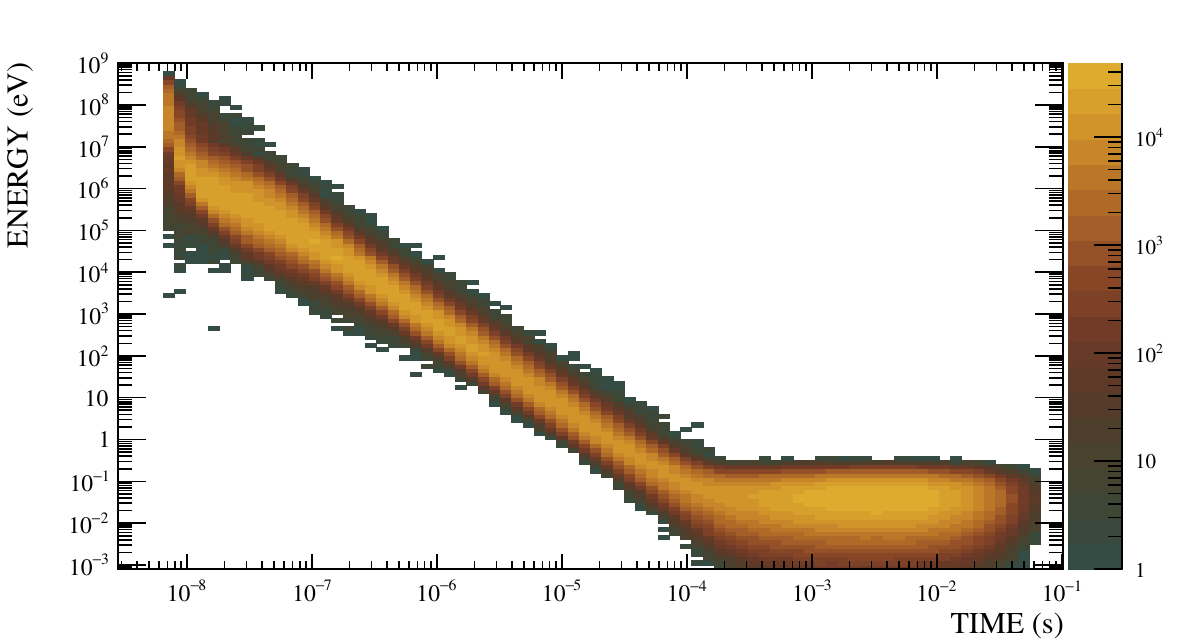}
	\caption{Energy evolution of neutrons crossing the region of the ion beam pipe. The neutron source was a 800-MeV spallation source in the middle of the proton beam pipe surrounded by a graphite cube of 300~cm length. High energy neutrons reach the ion pipe only at very early times. The later, the lower the neutron energy when crossing the ion pipe. Between 0.1~ms and 0.1~s, only thermal neutrons are present in the ion beam pipe. Both axes have a logarithmic binning with 10 bins/decade.}
	\label{fig:hockey_run_0171}
\end{figure}

\begin{figure}[H]
	\includegraphics[width=0.9\linewidth]{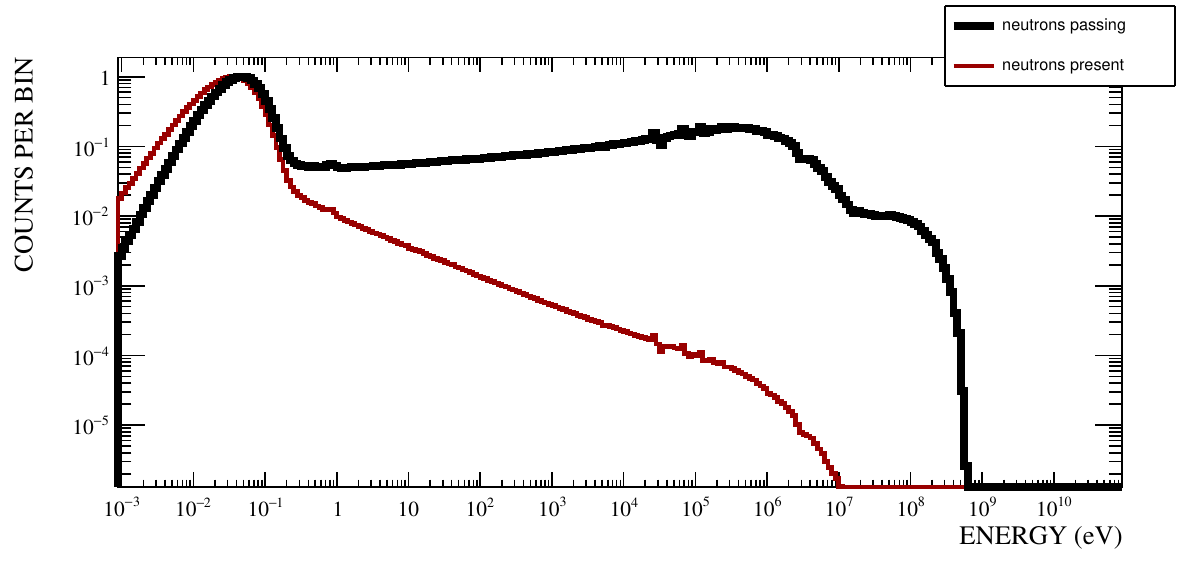}
	\caption{Black: A projection of Figure~\ref{fig:hockey_run_0171} onto the energy axis. This spectrum corresponds to netrons passing the ion pipe. Red: As black, but time-weighted. The time which in the neutrons spend in the ion pipe is used for this time-weighted neutron spectrum. Since slower neutron need much more time crossing the ion pipe, much more of them are present at any given time. This spectrum corresponds therefore to the neutrons actually present in the ion pipe. The x-axis has a logarithmic binning with 20 bins/decade.}
	\label{fig:neutron_spectra_ion_pipe_run_0171}
\end{figure}

In order to serve as a neutron target, a spallation source must be surrounded by an extended moderator. This moderator serves two key purposes: first, it traps neutrons allowing them significant time scattering within the moderator, which increases the neutron density and secondly, it slows them down producing an almost pure thermal spectrum, Figures~\ref{fig:hockey_run_0171} and~\ref{fig:neutron_spectra_ion_pipe_run_0171}. This thermalization enables the determination of the reaction energy in the center-of-mass directly from the ion beam energy. The optimal moderator geometry and material depends strongly on both the neutron spectrum of the spallation source and the specific experimental constraints \cite{RBD18}. Both aspects, thermalization and trapping, are extremely important for the success of this experimental approach.

The initial phase of this effort focuses on developing a proof-of-concept facility known as the Neutron Target Demonstrator (NTD), located at the Los Alamos Neutron Science Center (LANSCE) \cite{LBR90}. The goal of the NTD is to experimentally demonstrate the principle of neutron-induced reactions in inverse kinematics as simply as possible. Instead of a multi-pass ion storage ring, the NTD will employ a stable ion beam that passes through a graphite moderator containing a centrally located spallation target for the proton beam from LANSCE, oriented perpendicular to the ion beam. In addition, rather than use real-time reaction diagnostics the NTD will measure reaction products using off-line gamma-ray counting in a low background environment. This configuration enables an experimental test of the neutron target concept while minimizing all other technical challenges and requiring only minor adjustments to the experimental facilities at LANSCE \cite{CMR24}. 

To this end, a graphite moderator of approximately one cubic meter was assembled in order to serve as the moderator for the NTD. The purpose of the present work is to simulate and experimentally benchmark Monte Carlo simulations of the moderator performance and resulting neutron target properties in advance of the upcoming experimental campaign at LANSCE. Experiments were conducted at the Nuclear Science Laboratory (NSL) at the University of Notre Dame \cite{ISNAP} and the Cyclotron Institute at Texas A\&M University \cite{CI-TAMU} in order to assess moderator performance at a range of neutron energies relevant for the LANSCE spallation neutron source driven by an 800-MeV proton LINAC.

\section{Monte Carlo Simulations}

All simulations shown here were based on Geant-3.21 \cite{GEA93} including the gcalor package.

About 1~m$^3$ of graphite is available at LANSCE. This graphite consists of small blocks, which can be stacked into different shapes. It has been used in the past for neutron moderation. Table~\ref{tab:pieces} gives a summary of the graphite pieces used during the first experiments and the accompanying simulations.  

\begin{table}[h]
\begin{center}
  \caption{List of graphite moderator pieces used in simulations and during the first experiments, see Fig.~\ref{fig:ND_geometry}.}
  \label{tab:pieces}
    \begin{tabular}{ l | c | c}
    \hline
    \bf{ID}     & Geometry (cm)  \\ \hline
    \bf{A}      & 76.3x10.8x10.8     \\
    \bf{B}      & 16.8x10.8x10.8     \\
    \hline
  \end{tabular}
\end{center}
\end{table}

\begin{figure}[H]
	\includegraphics[width=0.45\linewidth]{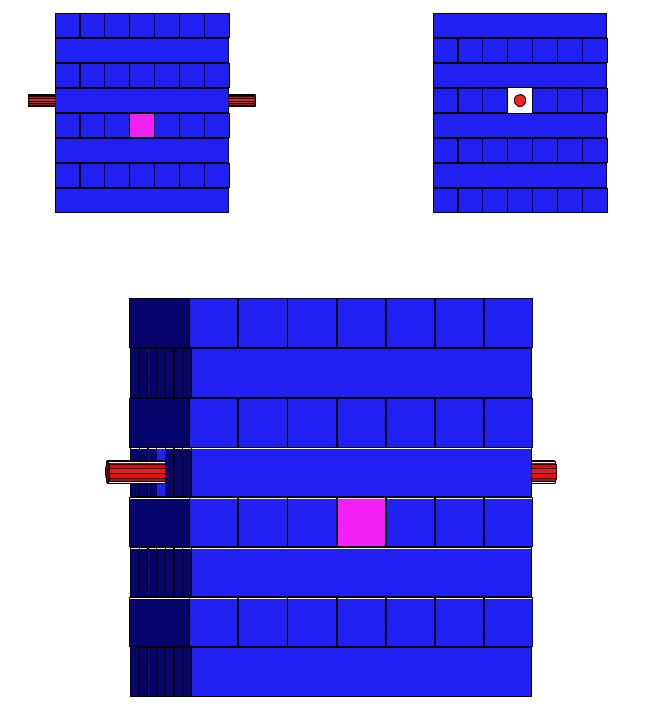}
	\includegraphics[width=0.45\linewidth]{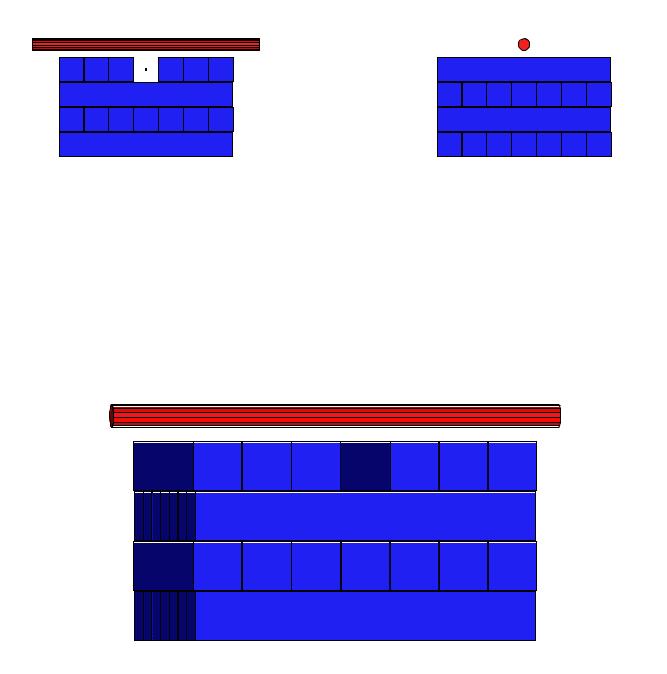}
	\caption{Simulated setup: Full cube on the left and half cube on the right. The red pipe indicates the position of the ion beam pipe, which corresponds to the position of the gold wires. The full cube consists of 54 large graphite blocks of type A (blue) and 2 smaller blocks of type B to cover the end of the proton beam line (pink). The half cube consists of only 27 pieces of type A.}
	\label{fig:ND_geometry}
\end{figure}

We determined the density of the graphite moderator material to be 1.62~g/cm$^3$ \cite{CRC24}. This value corresponds to reactor-grade graphite densities from the beginning of the nuclear industry \cite{AWB22}. Eight layers of graphite blocks were stacked on top of each other for the full cube. Six of those layers consisted of 7 blocks of type A. The two layers in the center of the cube had only 6 blocks. The central block in each of those layer was left out for proton and ion beam pipe, Fig.~\ref{fig:ND_geometry} (left). The downstream side of the proton beam line was plugged with 2 pieces of type B to prevent the neutrons from escaping at 0$^{\circ}$. While the full cube resembles a moderator that could actually be used as a real neutron target, the simulations can be stringently tested with only half of the assembly. If the top half is removed, the number of moderated neutrons passing the ion pipe will be drastically reduced while the number of high-energy neutrons will remain mostly unchanged, because they reach the wires largely without interacting with the graphite. For this purpose only, we made simulations and experiments also with a half cube. Only the bottom 4 layers of the full cube were used for the half cube. In addition, the plugs were removed, Fig.~\ref{fig:ND_geometry} (right).  

\section{Activation method}\label{sec:activation}

We applied the activation method to compare the simulated time-integrated neutron flux inside the moderator with experimental data. For that purpose, thin pieces of gold wire were placed along entire length of the future beam pipe on sturdy, low-density package foam, (Fig.~\ref{fig:ND_geometry}). That foam was not included in the simulations. The gold wire was cut into pieces of approximately 2~cm (about 20~mg) in order to determine the activation of the gold as a function of position inside the beam pipe. 

Gold has one stable isotope, $^{197}$Au. After neutron capture, $^{198}$Au is formed, which conveniently decays with 2.7~d of half-live emitting an easily detectable $\gamma$-ray with 412~keV. High-energy neutrons produce also neutron-deficient gold isotopes via neutron-removal reactions such as (n,2n), (n,3n), (n,4n), see Table~\ref{tab:au-reactions}, but those reaction channels were only open for the last two runs at the highest energy reported here.


\begin{table}[h]
\begin{center}
  \caption{List of neutron-induced reactions on $^{197}$Au. The neutron capture cross section is a standard over a wide energy range.}
  \label{tab:au-reactions}
    \begin{tabular}{ l |p{1.4cm}| c | c |c }
    \hline
    \bf{Reaction}  & $E_{min}$ (MeV)   & Product & $t_{1/2}$  & $\gamma$-rays (keV)  \\ \hline
    (n,$\gamma$) & 0.0 & $^{198}$Au  & 2.7 d & 412     \\
    (n,2n) & 8.072 & $^{196}$Au  & 6.2 d & 333, 356, 426      \\
    (n,3n) & 14.72 & $^{195}$Au  & 186 d & 99      \\
    (n,4n) & 23.14 & $^{194}$Au  & 38 h & 294, 328, 1469      \\
    \hline
  \end{tabular}
\end{center}
\end{table}

Post irradiation, the gold wires were separately placed in front of calibrated High-Purity Germanium (HPGe) detectors. The most important line used to analyze the gold activity is the 412-keV line emitted during the decay of $^{198}$Au. The parameters including uncertainties of all lines used for the analysis reported here are summarized in Table~\ref{tab:au-decays}.

\begin{table}[h]
\begin{center}
  \caption{List of decay parameters used during the analysis. The decay of $^{7}$Be was used to determine the number of neutrons produced via the $^{7}$Li(p,n)$^{7}$Be reaction. }
  \label{tab:au-decays}
    \begin{tabular}{ l |c | c | c |r}
    \hline
    Isotope     & $t_{1/2}$     & $\gamma$-rays (keV) & $I_\gamma$ & Ref.\\ \hline
    $^{194}$Au  & 38.06 h       & 328.470(6)       & 0.628 & \cite{ChS21} \\
    $^{196}$Au  & 6.1669(6) d   & 355.73(5)        & 0.87   & \cite{Xia07} \\
    $^{198}$Au  & 2.6941(2) d   & 411.80205(17)    & 0.9562 & \cite{XiM16} \\
    $^{7}$Be    & 53.22(6) d    & 477.6035(20)     & 0.1044 & \cite{TCG02} \\
    \hline
  \end{tabular}
\end{center}
\end{table}

We used the following equations to determine the number of produced unstable atoms \cite{REF18}:

\begin{equation}
  N_{product} = \frac{C_\gamma}{\epsilon_\gamma I_\gamma} \frac{1}{f_m f_b f_w}
\end{equation}

with

\begin{itemize}
  \item $N_{product}$ the number of produced atoms (e.g. $^{198}$Au).
  \item $C_\gamma$ the number of counts in the spectrum for the corresponding line (e.g. 412 keV).
  \item $\epsilon_\gamma$ the photo peak detection efficiency for the corresponding line, depends on the setup.
  \item $I_\gamma$ the emission intensity of the corresponding photon.
  \item $f_m$ correction for finite $\gamma$-counting measurement time: 
        \begin{equation}
           f_m = 1-\mathrm{e}^{-\lambda t_m}
        \end{equation}
        with the measuring time $t_m$ and the decay constant $\lambda=\ln(2)/t_{1/2}$ .
  \item $f_w$ correction for number of decays during the time $t_w$ between end of activation and beginning of $\gamma$-counting: 
        \begin{equation}
           f_w = \mathrm{e}^{-\lambda t_w}
        \end{equation}
  \item $f_b$ correction for number of decays during the activation time $t_a$. If the neutron flux is constant, this correction is simply:
        \begin{equation}
           f_b = \frac{\lambda t_a}{1-\mathrm{e}^{-\lambda t_a}}
        \end{equation}
  If the neutron flux varies and is monitored with constant intervals, it can be expressed as: 
        \begin{equation}
           f_b = \frac{\sum_i \Phi_i \mathrm{e}^{-\lambda (t_a - t_i)}}{\sum_i \Phi_i}
        \end{equation}
        with $\Phi_i$ the measured neutron flux at time $t_i$ after beginning of the irradiation.
\end{itemize}

For the comparison of experimental data with simulations, it is useful to define the yield as the ratio of produced radioactive atoms to the number of sample atoms and neutrons \textit{emitted} by the source:

\begin{equation}
  Y = \frac{N_{product}}{N_{sample} N_{neutron}}
\end{equation}

All results discussed in section \ref{sec:measurements} will refer to $Y$. 

\section{Measurements}\label{sec:measurements}

We performed a series of activation experiments with different neutron energy distributions and moderator setups. The experiments were carried out at the Nuclear Science Laboratory (NSL) at the University of Notre Dame, IN, USA \cite{ISNAP} and the Cyclotron Institute (CI) at Texas A\&M University \cite{CI-TAMU}. The parameters of the different experiments are summarized in Table~\ref{tab:irradiations}. The idea was to verify the simulations first with rather low-energy neutrons and tailored spectra before performing the final experiment at LANSCE with a 800-MeV spallation source.

\begin{table}[h]
\begin{center}
  \caption{List of irradiations. The configuration includes the number of pieces of type \textbf{A} and \textbf{B} as defined in Table~\ref{tab:pieces} $t_A$ is the activation time.}
  \label{tab:irradiations}
    \begin{tabular}{ c |c| c | c |c |c }
    \hline
    \bf{ID}  & $E_{proton}$  & Reaction       & configuration           & $t_A$ & facility \\ 
             & (MeV)         &                & (\textbf{A}/\textbf{B}) & (h) & facility \\ \hline
    I-a      & 1.95          & $^{7}$Li(p,n)  & full (54/2)             & 2  & NSL, Notre Dame \\
    I-b      & 1.95          & $^{7}$Li(p,n)  & half (27/0)             & 6  & NSL, Notre Dame \\
    II-a     & 2.5           & $^{7}$Li(p,n)  & full (54/2)             & 2  & NSL, Notre Dame \\
    II-b     & 2.5           & $^{7}$Li(p,n)  & half (27/0)             & 5  & NSL, Notre Dame \\
    III-a    & 9.0           & $^{9}$Be(p,n)  & full (54/0)             & 22  & CI, Texas A\&M \\
    III-b    & 9.0           & $^{9}$Be(p,n)  & half (27/0)             & 21  & CI, Texas A\&M \\
    IV-a     & 45.0          & $^{9}$Be(p,n)  & full (54/0)             & 18  & CI, Texas A\&M \\
    IV-b     & 45.0          & $^{9}$Be(p,n)  & half (27/0)             & 18  & CI, Texas A\&M \\
    \hline
  \end{tabular}
\end{center}
\end{table}

\subsection{$\gamma$-detection}

\begin{figure}[H]
	\includegraphics[width=\linewidth]{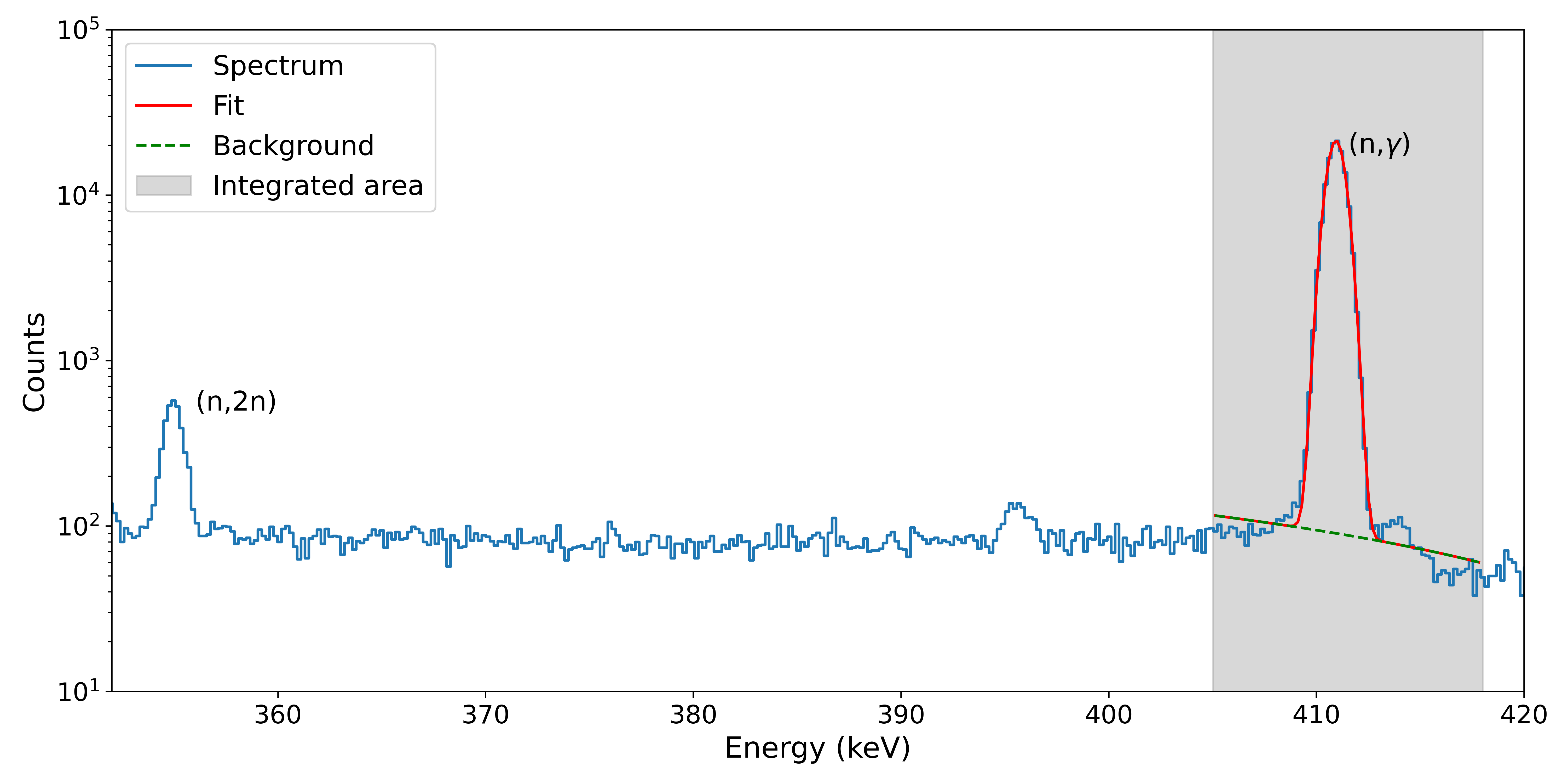}
	\caption{ Gamma spectrum of 4 activated gold foils, positioned in the middle of the half-moderator at 45 MeV, at Texas A$\&$M (run~IVb)). 
    }
	\label{fig:gamma_speclog}
\end{figure}

Following the activations, the gold foils were analyzed via $\gamma$-spectroscopy using various high-purity germanium (HPGe) detectors. For this reason, we made sure that the cross-calibration was done using at least one gold wire at both setups.
To determine the yield, it is essential to evaluate the ratio between the number of produced particles and the number of sample atoms. The number of sample atoms is calculated based on the weight of the gold wire. Determining the number of produced particles requires an analysis of the $\gamma$-spectra. Figure~\ref{fig:gamma_speclog} illustrates a representative fit of the 412~keV $\gamma$-line following the $\beta$-decay of $^{198}$Au. The fitting function used to model the peak and background is a Gaussian combined with a linear background. The $\gamma$-line was integrated to determine the total number of counts. The background contribution was subtracted using the parameters obtained from the best fit of the peak. Additionally the $\gamma$-line at 356~keV from $^{196}$Au is visible for the runs with 45~MeV proton energy (IV-a and IV-b).

In addition to the statistical uncertainties resulting from the $\gamma$-counting, we assumed an additional 2\% individual systematic uncertainty for each wire. This accounts for the uncertainty in the mass determination as well as small deviation in the efficiency resulting from slightly different shapes and position of each wire. The overall correlated systematic uncertainties resulting from $\gamma$-detection efficiencies, intensities and half-lives are only important for the runs at the NSL, where we were able to determine a absolute normalization (section \ref{sec:ND}).

\subsection{Neutron source: $^7$Li(p,n)}\label{sec:ND}

\begin{figure}[H]
	\includegraphics[width=0.9\linewidth]{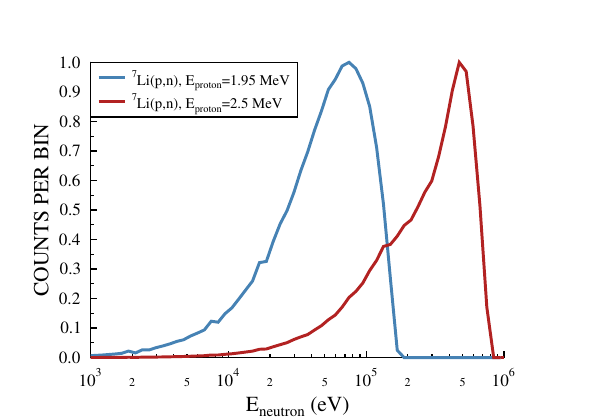}
	\caption{Source energy distribution of neutrons produced inside the moderator. The effect of the moderator on the neutron distribution is not included here. All spectra are normalized to maximum 1. The binning of the x-axis is logarithmic. The neutron source reaction was $^{7}$Li(p,n) using a 0.1 mm Li layer. The spectra are based on the PINO code \cite{RHK09,pino_online}.}
	\label{fig:ND_spectra}
\end{figure}

The first four runs (Ia, Ib, IIa, IIb) were performed at the NSL at the University of Notre Dame, IN, USA \cite{ISNAP}. The current on target was 10~$\mu$A for runs Ia, IIa and IIb and 20~$\mu$A for Ib. We used one single metallic lithium target with 100~$\mu$ thickness evaporated onto a copper backing for all runs. This target was thick enough to slow down protons with 1.95~MeV as well as 2.5~MeV below the neutron production threshold \cite{RHK09,pino_online}. We checked the neutron spectrum before and after each irradiation using a $^3$He detector while the moderator was removed. These measurements showed no changes in particular at the lower end, hence the lithium target was always thick enough to slow down the protons below the neutron production threshold. We used the simulation tool PINO \cite{RHK09,pino_online} as a double-differential neutron generator. Figure~\ref{fig:ND_spectra} shows the resulting source energy distribution.

The $^7$Be activity was measured after each run with a CeBr$_3$ detector without removing the lithium target. The efficiency of this detector was not absolutely calibrated, but we made sure that the position relative to the lithium target was reproduced for each measurement. These measurements were only intended to determine the \textit{relative} neutron flux after each of the four runs with high precision. 

\begin{figure}[H]
	\includegraphics[width=0.9\linewidth]{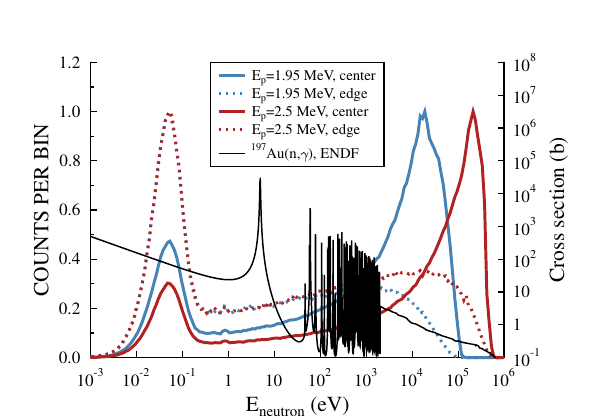}
	\caption{Simulated energy distribution of neutrons inside the in the region of the gold wires resulting from the source distribution shown in Figure~\ref{fig:ND_spectra} in the central region as well as close to the edge of the moderator. Each spectrum is separately normalized to have its maximum value at 1. In addition, the $^{197}$Au(n,$\gamma$) cross section used to determine the capture yield is shown \cite{BCC18}.}
	\label{fig:ND_spectra_pipe}
\end{figure}

Four examples of the simulated neutron spectra at the position of the gold wires are shown in Figure~\ref{fig:ND_spectra_pipe}. For each of the two proton energies, the spectrum in the center as well as close to the edge of the moderator is shown. The most striking difference is the presence of completely moderated, thermal neutrons around 25~meV. The central spectra predict much less moderation than the ones at the edge. The obvious reason is that much less moderator material is between this position and the neutron production target. The same figure also contains the $^{197}$Au(n,$\gamma$) cross section (ENDF/B-VIII.0, \cite{BCC18}) we used to estimate the capture yield by folding the spectrum with the cross section. 

\begin{figure}[H]
	\includegraphics[width=0.9\linewidth]{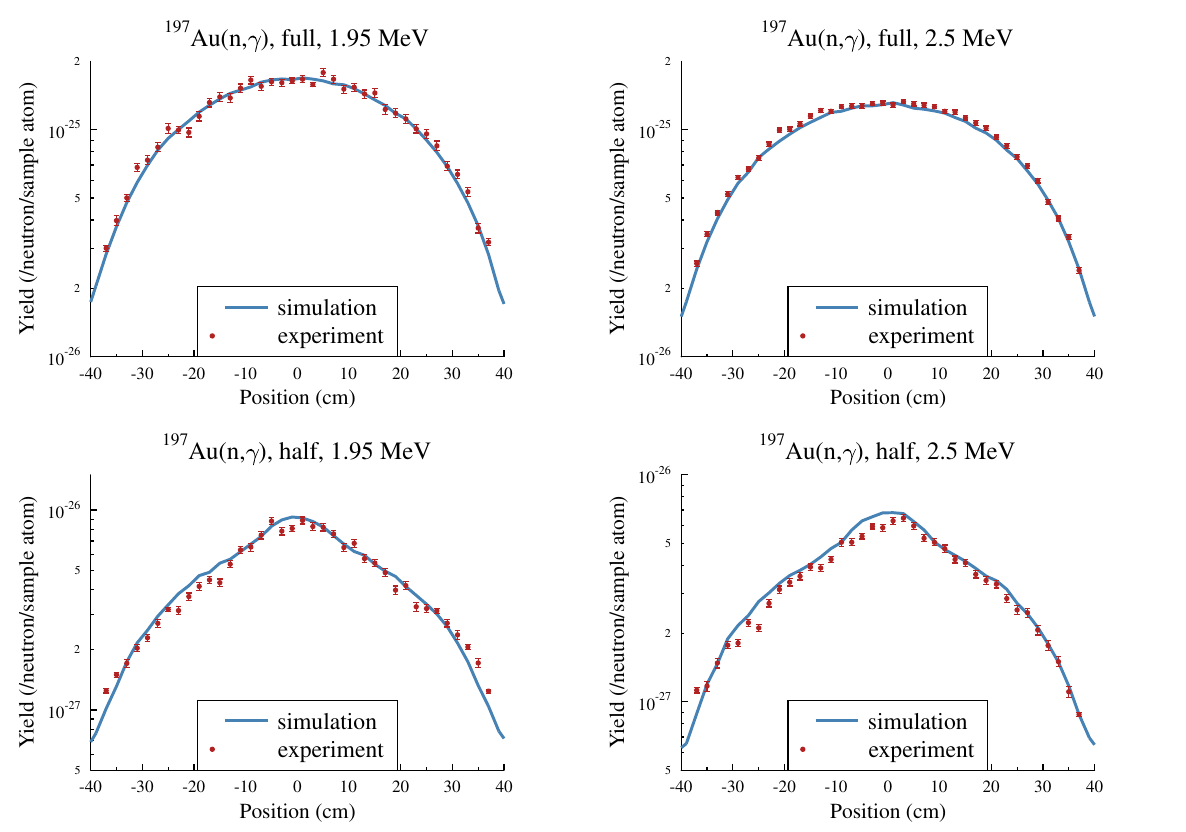}
	\caption{Results for all four runs performed at the NSL. Each plot shows the position inside the moderator, where 0 is the center, just above the lithium target. All four plots have one common normalization factor - the total number of neutrons.}
	\label{fig:ND_results}
\end{figure}

Figure~\ref{fig:ND_results} shows a comparison between simulated and experimental yields for all four runs. It is important to emphasize that only one single normalization parameter was used in this plot for the experimental data - the total number of neutrons. The uncertainties of the experimental data contain the statistical as well as the 2\% individual systematic uncertainties. The agreement between experiment and simulation over the entire range of the moderator and one order of magnitude is remarkable.

After 6 months, we determined the total $^7$Be activity with a calibrated HPGe detector. The agreement between the number of neutrons derived from the $^7$Be activity and the gold activities (Figure~\ref{fig:ND_results}) was within estimated 5\% uncertainty of the determination of the total neutron flux.

\subsection{Neutron source: $^9$Be(p,n)}

\begin{figure}[H]
	\includegraphics[width=0.9\linewidth]{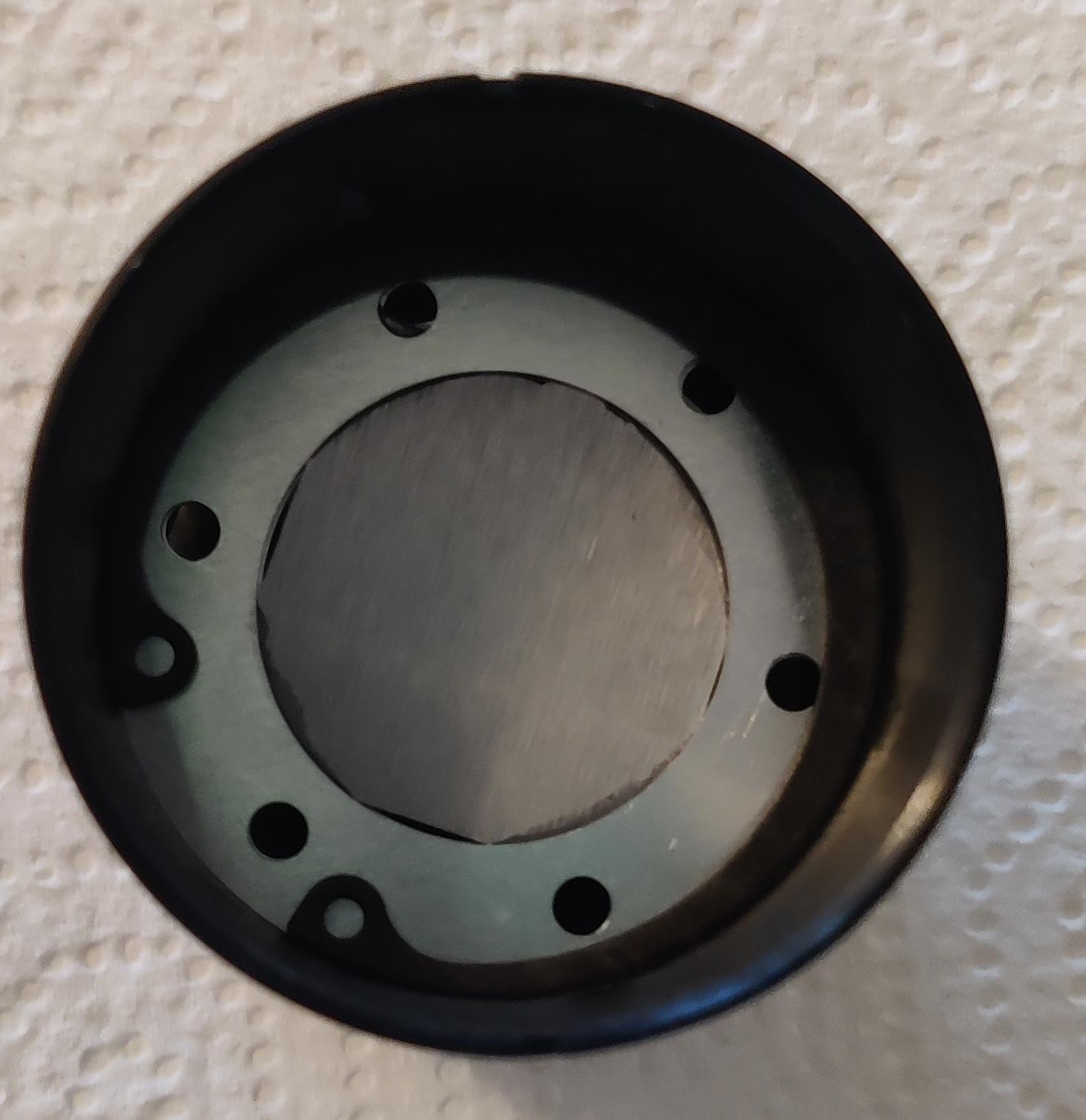}
	\caption{Beryllium target used at the Texas A$\&$M cyclotron. The 0.5~mm Be-foil was glued to an aluminum frame, which was mounted inside a target holder (black can). The Al-frame contains holes, which allows evacuating and venting of the beam pipe from one side. The target holder was then slid into the beam and held in place with 2 snap rings.}
	\label{fig:be_target}
\end{figure}

The experiment at the Texas A$\&$M Cyclotron Institute \cite{CI-TAMU} was carried out using two proton beam energies: 9~MeV and 45~MeV. A 0.5~mm self-supporting beryllium (Be) foil served as the neutron source, facilitating the $^9$Be(p,n) reaction, see Fig.~\ref{fig:be_target}. This target is not thick enough to slow down the protons below the (p,n) threshold, but target degradation and hence changes of the neutron spectrum is not an issue for this target, since beryllium can stand much higher temperatures than metallic lithium. The current on target was 1~$\mu$A for the 9~MeV runs (IIIa, IVb) and 200~nA for the 45~MeV runs (IVa, IVb).

\begin{figure}[H]
	\includegraphics[width=0.9\linewidth]{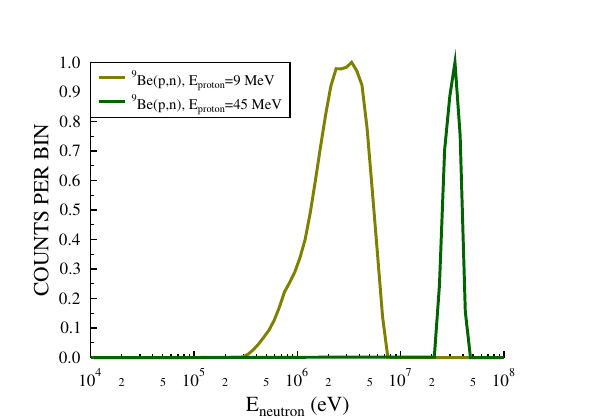}
	\caption{Source energy distribution of neutrons produced inside the moderator. The effect of the moderator on the neutron distribution is not included here. All spectra are normalized to maximum 1. The binning of the x-axis is logarithmic. The neutron source reaction was $^{9}$Be(p,n) using a self-supporting 0.5~mm Be foil. The spectra are based on the PINO code \cite{RHK09,pino_online}. }
	\label{fig:TAM_spectra}
\end{figure}

Similarly to the runs at NSL, we used the simulation tool PINO \cite{RHK09,pino_online} to generate double-differential neutron spectra. The expected neutron source spectra are shown in Figure~\ref{fig:TAM_spectra}. For each proton energy, measurements were taken using both a full (IIIa, IVa) and a half moderator (IIIb, IVb) cube configuration, resulting in a total of four activation measurements. 

\begin{figure}[H]
	\includegraphics[width=0.9\linewidth]{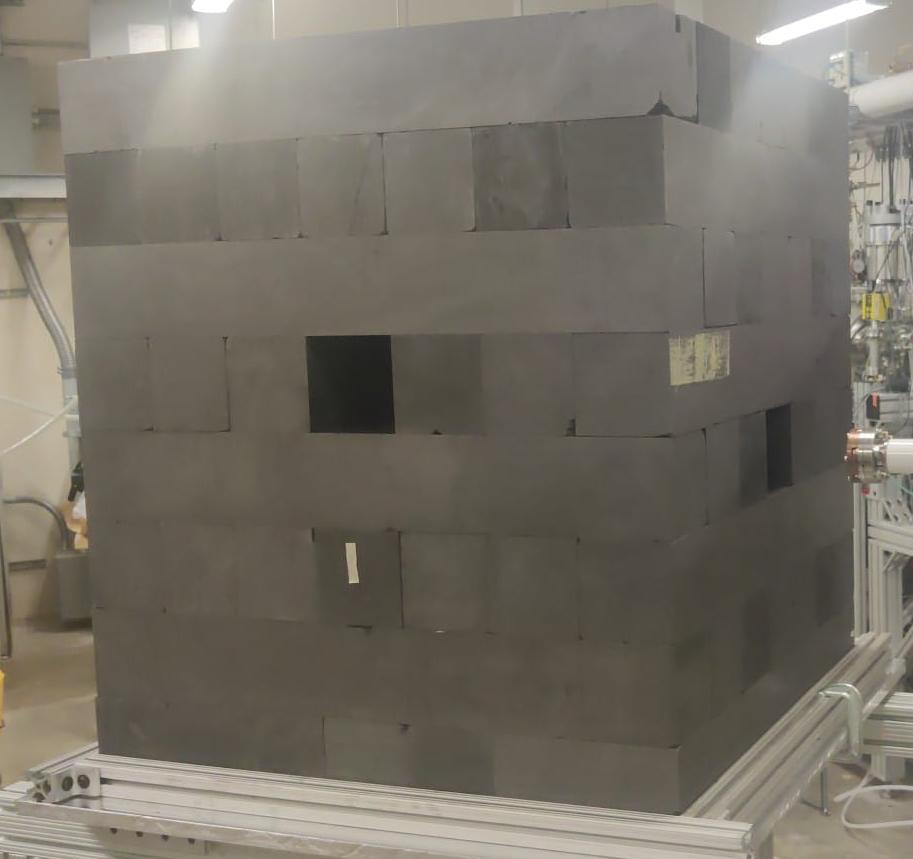}
	\includegraphics[width=0.9\linewidth]{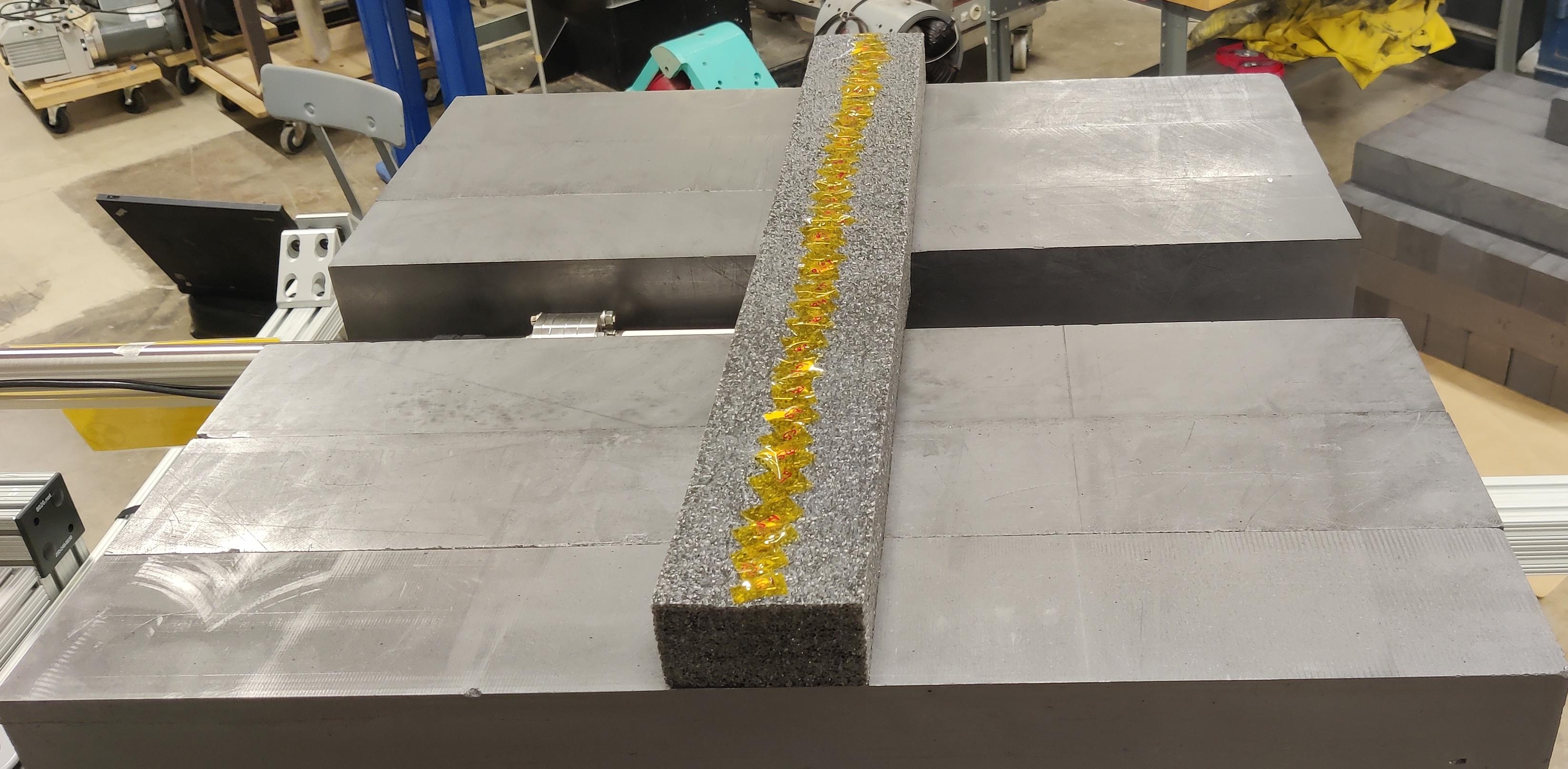}
	\caption{Top: View of the full cube stacked with of 54 graphite blocks. Bottom: View of the half-cube assembly including the gold wires used to measure the neutron flux via activation. Each gold foil was counted separately after the irradiation.}
	\label{fig:half_cube_tamu}
\end{figure}

In each setup, gold foils were placed within the moderator at the intended ion beam pipe location, see Fig.~\ref{fig:half_cube_tamu}. 

\begin{figure}[H]
	\includegraphics[width=0.9\linewidth]{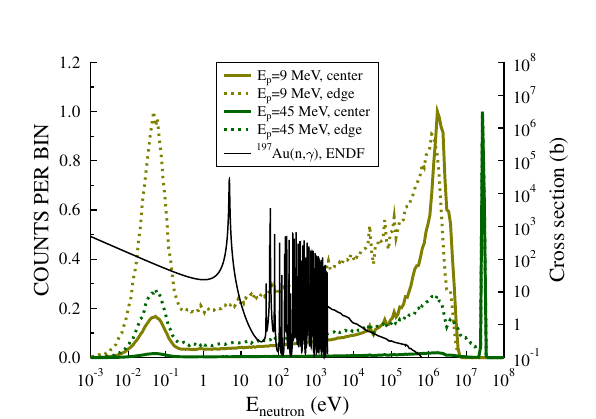}
	\caption{Simulated energy distribution of neutrons inside the in the region of the gold wires resulting from the source distribution shown in Figure~\ref{fig:TAM_spectra} in the central region as well as close to the edge of the moderator. Each spectrum is separately normalized to have its maximum value at 1. In addition, the $^{197}$Au(n,$\gamma$) cross section used to determine the capture yield is shown \cite{BCC18}.}
	\label{fig:TAM_spectra_pipe}
\end{figure}

Four examples of the simulated neutron spectra at the position of the gold wires are shown in Figure~\ref{fig:TAM_spectra_pipe}. For each of the two proton energies, the spectrum in the center as well as close to the edge of the moderator is shown. Again, the central spectra predict much less moderation than the ones at the edge and in general the fraction of thermal neutron is less for the runs with lower proton energies. The same figure also contains the $^{197}$Au(n,$\gamma$) cross section (ENDF/B-VIII.0, \cite{BCC18}) we used to estimate the capture yield by folding the spectrum with the cross section. We used the ENDF/B-VIII.0 library also for the (n,2n) channel.

\subsubsection{$E_{proton}=9$ MeV}

\begin{figure}[H]
	\includegraphics[width=0.9\linewidth]{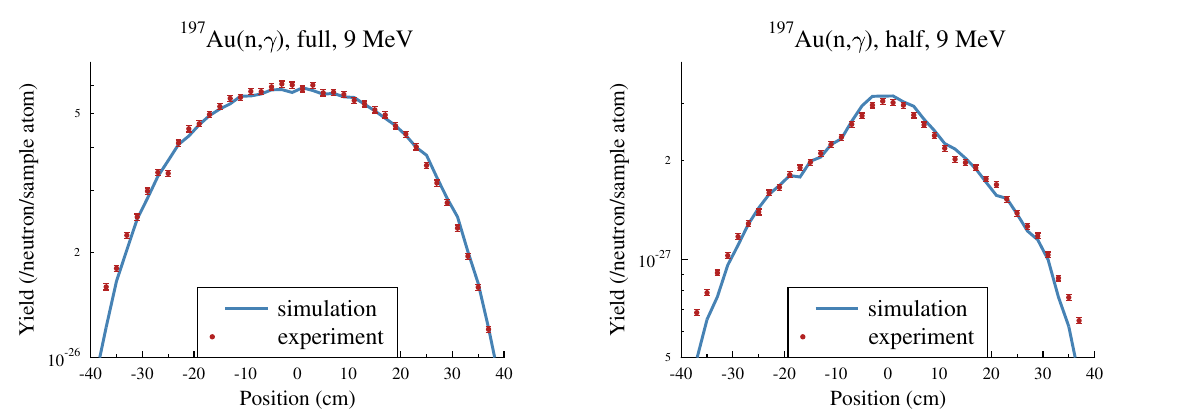}
	\caption{Results for the 9~MeV runs performed at Texas A\&M. Each plot shows the position inside the moderator, where 0 is the center, just above the beryllium target. Each experimental data set (half and full cube) is separately normalized to the simulations.}
	\label{fig:TAM_9_MeV}
\end{figure}

Figure~\ref{fig:TAM_9_MeV} shows a comparison between simulated and experimental yields for the runs with 9~MeV proton energy (IIIa, IIIb). Since we could not measure the relative production of neutrons in each run, each experimental data set is separately normalized to the simulated data. The uncertainties of the experimental data contain again the statistical as well as the 2\% individual systematic uncertainties. The agreement between experiment and simulation for the full cube is again excellent. The half cube shows slight deviations at the wings, which we attribute to room background. This feature is also visible in the 45~MeV run. The room background has very likely two causes - neutrons produced at the beryllium target and scattered back from the walls and neutrons produced by protons hitting other beam line components.

\subsubsection{$E_{proton}=45$ MeV}

\begin{figure}[H]
	\includegraphics[width=0.9\linewidth]{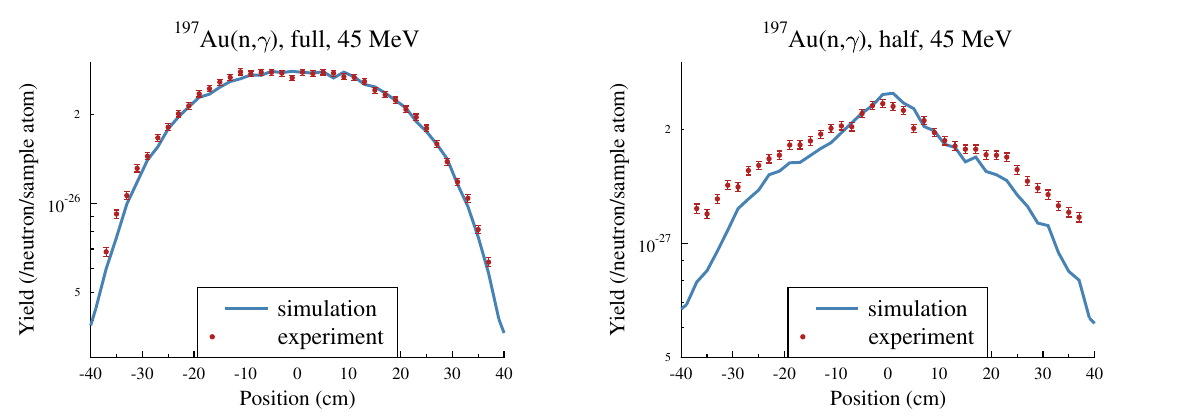}
	\caption{Results for the $^{198}$Au activity after the 45~MeV runs performed at Texas A\&M. Each plot shows the position inside the moderator, where 0 is the center, just above the beryllium target. Each experimental data set (half and full cube) is separately normalized to the simulations.}
	\label{fig:TAM_45_MeV_ng}
\end{figure}

The main result of this run is shown in Figure~\ref{fig:TAM_45_MeV_ng}, which illustrates a comparison between simulated and experimental neutron capture yields for the runs with 45~MeV proton energy (IVa, IVb). Since we could not measure the relative production of neutrons in each run, each experimental data set is separately normalized to the simulated data. The uncertainties of the experimental data contain again the statistical as well as the 2\% individual systematic uncertainties. The agreement between experiment and simulation for the full cube is again excellent. The half cube shows significant deviations at the wings, which we again attribute to the significant room background. The graphite moderator acts as a shield against those neutrons in case of the full cube.

\begin{figure}[H]
	\includegraphics[width=0.9\linewidth]{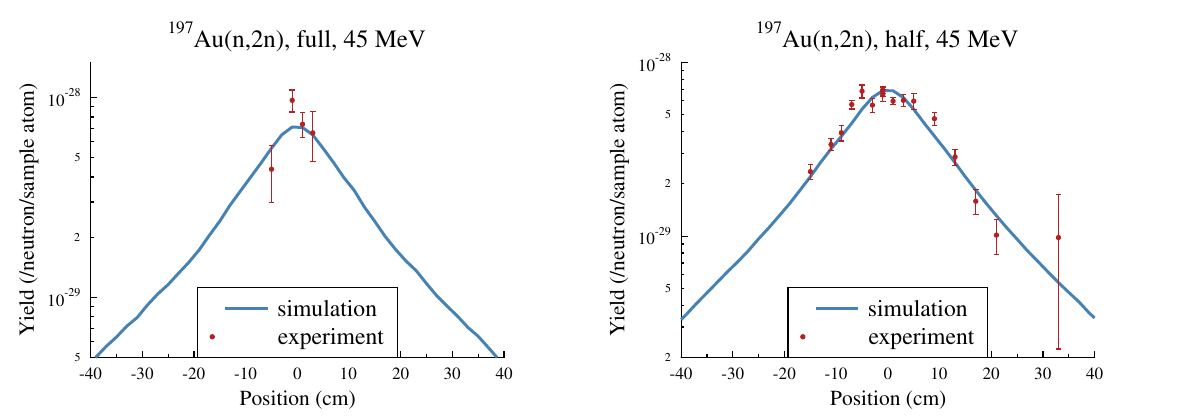}
	\caption{Results for the $^{196}$Au activity after the 45~MeV runs performed at Texas A\&M. Each plot shows the position inside the moderator, where 0 is the center, just above the beryllium target. Each experimental data set (half and full cube) is separately normalized to the simulations.}
	\label{fig:TAM_45_MeV_n2n}
\end{figure}

Figure~\ref{fig:TAM_45_MeV_n2n} shows a side effect, the comparison between simulated and experimental (n,2n) yields for the same runs. Again, each experimental data set is separately normalized to the simulated data and the uncertainties of the experimental data contain again the statistical as well as the 2\% individual systematic uncertainties. The agreement between experiment and simulation is in both cases is again excellent. This is important because one would not expect much of an impact of the moderator on the corresponding high-energy neutrons. Full and half cube show virtually the same spacial distribution in the simulations as well for the wires where a measurement was possible. 

We attempted to detect the activity from the (n,3n) channel, but even 8 months after the irradiation it was not visible above the room background of the setups available. We could see the rather short-lived (n,4n) activity, but not with enough statistics for a useful analysis. We expect to see these channels more pronounced during high-energy runs with 800~MeV protons.

\section{Conclusions}\label{sec:conclusions}

\begin{figure}[H]
	\includegraphics[width=0.9\linewidth]{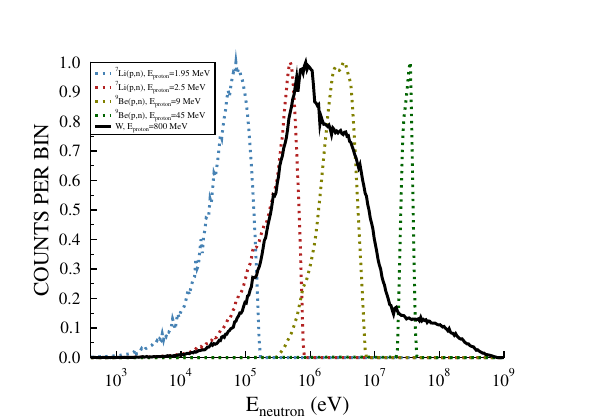}
	\caption{Neutron source spectra during activations described in this report ($^7$Li(p,n) and $^9$Be(p,n)) and the planned experiment at the LANSCE facility using spallation on tungsten triggered by 800~MeV protons. The x-axis has logarithmic binning.}
	\label{fig:ALL_spectra}
\end{figure}

The experiments presented in here covered the central part of the spallation neutron energy distribution, Figure~\ref{fig:ALL_spectra}. We found excellent agreement between simulations and experiment for the spatial distribution of neutrons inside the full graphite cube. 
At lower proton energies (2.5 MeV and 5 MeV) we could also confirm the simulations. At higher proton energies (9 MeV and 45 MeV), our measurements of the (n,$\gamma$) channel were disturbed by room background affecting the low-activity wings of the distribution. The high-energy (n,2n) channel was not affected by this disturbance and we found again very good agreement.

The different neutron production mechanisms used in this study can be directly compared with spallation as the neutron source. The results of this comparison for a nominal beam current of 10~$\mu$A are summarized in Table~\ref{tab:n-densities}. Each of the 3 targets (Li, Be, W) are thick enough that incident protons will be slowed down below the neutron production threshold and as such the neutron yield is at its maximum. As outlined in \cite{RGH17}, the conversion from the sum of all times neutrons spend in the ion beam pipe into an areal neutron density of the neutron target as seen by the passing ion is linear. Each proton produces a certain amount of neutrons. These neutrons travel through the setup, are moderated and eventually pass the ion beam pipe. The proton-averaged total time period $t_{\mathrm{neutron}}$ sums up the time periods that all neutrons spend inside the beam pipe per incident proton (see Table~\ref{tab:n-densities}). To estimate the average number of neutrons inside the beam tube, $\bar{n}_{\mathrm{neutron}}$, we multiply $t_{\mathrm{neutron}}$ by the number of protons hitting the neutron production target per time (proton current/elementary charge):

\begin{equation}\label{eq:neutrons_in_target}
  \bar{n}_{\mathrm{neutron}} = \frac{I_{\mathrm{proton}}}{e} t_{\mathrm{neutron}}
\end{equation}

The corresponding areal neutron density $\eta_{\mathrm{A}}$ is calculated by dividing $\bar{n}_{\mathrm{neutron}}$ by the cross sectional area of the ion beam pipe $A_{\mathrm{ion\;pipe}}$:

\begin{equation}\label{eq:neutrons_in_target_areal}
  \eta_{\mathrm{neutron}} = \frac{\bar{n}_{\mathrm{neutron}}}{A_{\mathrm{ion\;pipe}}} = \frac{I_{\mathrm{proton}}}{e} \frac{t_{\mathrm{neutron}}}{A_{\mathrm{ion\;pipe}}}
\end{equation}

\begin{table}[h]
\begin{center}
  \caption{Comparison of neutron densities for different production mechanisms and a nominal proton current of 10~$\mu$A. At LANSCE, 800~MeV proton beam is used for most experiments. However, also a less favorable 100~MeV line is available and we discussed this possibility in \cite{RCO25}. The LANSCE accelerator in combination with the Proton Storage Ring (PSR) is capable of delivering beams over a wide range of parameters, details can be found in \cite{LBR90, ZMK18}.}
  \label{tab:n-densities}
    \begin{tabular}{ l | c |  c | c}
    \hline
    production    & $E_{proton}$   & $t_{neutron}$    & $\eta_A$ at 10~$\mu$A \\
    principle     & (MeV)          & ($\mu$s/proton)  & (10$^{3}$/cm$^2$)     \\ \hline
    $^7$Li(p,n)   & 2.5            & 0.0041           & 13       \\
    $^7$Li(p,n)   & 5              & 0.018            & 57      \\
    $^9$Be(p,n)   & 10             & 0.0089           & 28       \\
    $^9$Be(p,n)   & 50             & 0.033            & 100    \\
    W(spallation) & 100            & 0.26             & 830    \\
    W(spallation) & 800            & 12               & 38000    \\
    \hline
  \end{tabular}
\end{center}
\end{table}

The expected neutron target density for spallation neutrons driven by a 10~$\mu$A proton beam at 800~MeV is 3.8~10$^7$ n/cm$^2$, which will be sufficient to demonstrate the neutron target concept.

\section{Outlook}

\begin{figure}[H]
	\includegraphics[width=0.9\linewidth]{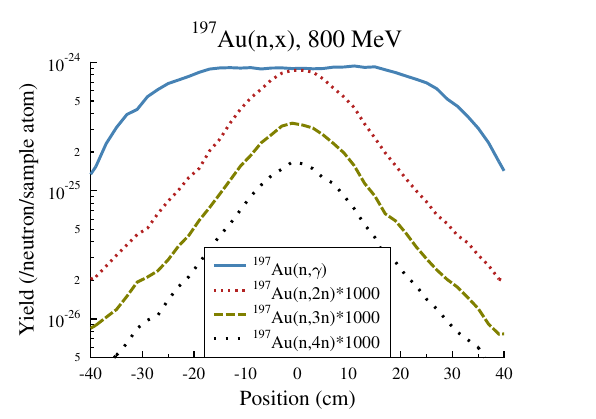}
	\caption{Expected (n,x) yields following the irradiation of a tungsten target surrounded by the full cube used in the experiments so far.}
	\label{fig:LANSCE}
\end{figure}

The next step in this project is a similar gold activation experiment at LANSCE to extend the current validation effort to the full energy regime produced by a spallation neutron source \cite {RCC23}. Figure~\ref{fig:LANSCE} shows the simulated spatial flux distributions for the various neutron-induced reactions on gold in the planned spallation experiment. The simulation assumes the full moderator cube without plugs and models the neutron source as a tungsten cylinder 7.5 cm in diameter and 25 cm in length. This spallation target already exists at LANSCE and has previously been used in a similar experimental setup \cite{RHO05}. This measurement will be followed by an inverse-kinematics experiment using a krypton beam from an ion source passing through the spallation neutron target in the moderator \cite{CMR24}. 

\section{Summary}

Within the framework of the Neutron Target Demonstrator project at LANL, we measured the neutron flux distribution inside a graphite moderator for four different source neutron-energy distributions ranging from  1~keV to about 50 MeV. The volume of the full moderator was about 1~m$^3$. We found excellent agreement between simulation and experiment over the entire energy range for the full cube. We found very good agreement also for experiments using only half of the cube except for the neutron capture channel at high source energies.

\begin{figure}[H]
	\includegraphics[width=0.9\linewidth]{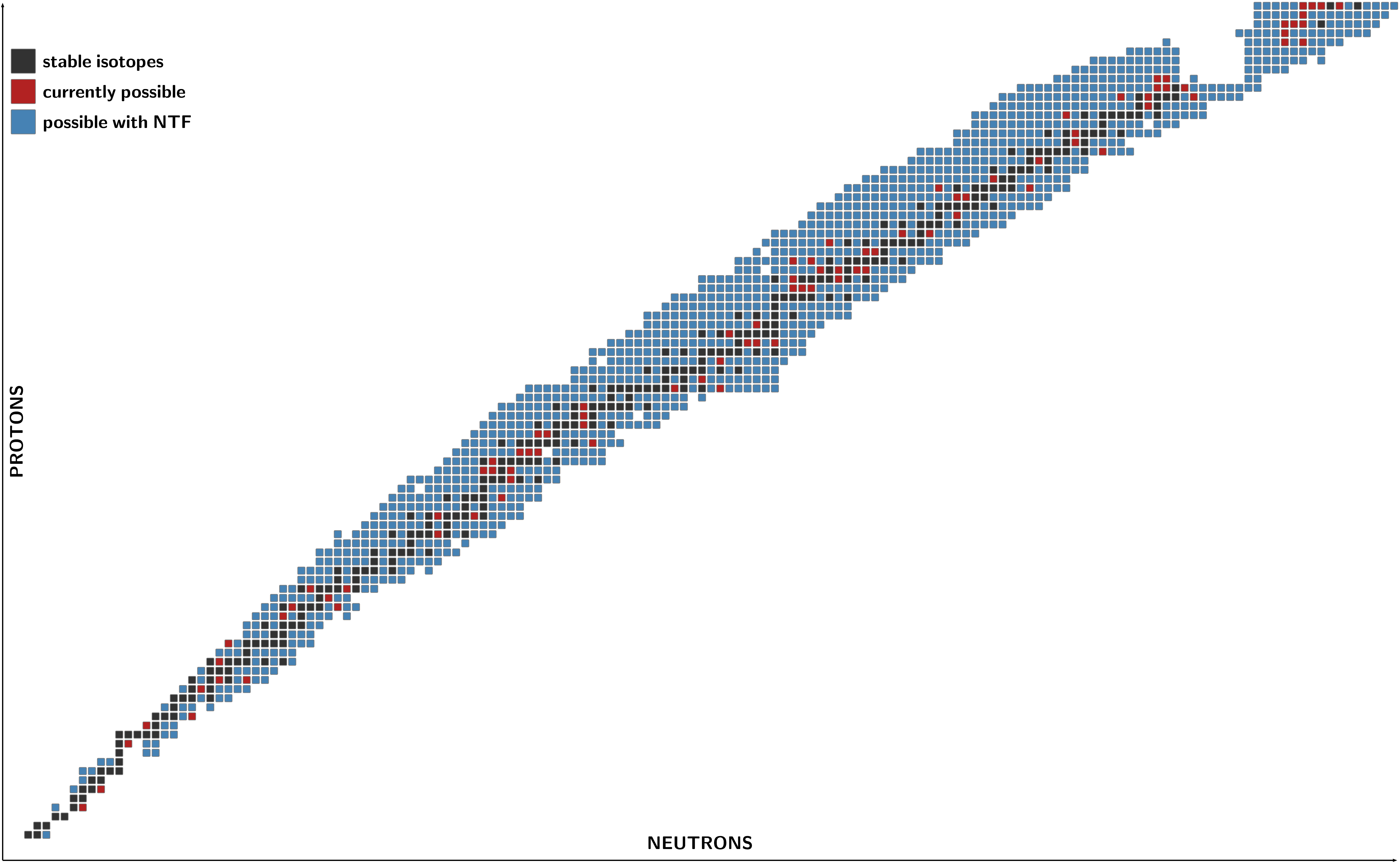}
	\caption{Chart of nuclides with stable isotopes (black), radioactive isotopes that can be investigated with current state-of-the-art instruments (red) and the additional isotopes that can be investigated at the future Neutron Target Facility (blue).}
	\label{fig:very_short_long_halflives}
\end{figure}

These initial neutron flux characterizations and neutron target tests provide a validated foundation for the development of a Neutron Target Facility (NTF) based on the principles outlined in \cite{RGH17}. If realized, the NTF would enable direct investigation of isotopes with half-lives as short as fractions of a minute, which would significantly expand the range of nuclei accessible for experimental study, as illustrated in Figure~\ref{fig:very_short_long_halflives}.

\section{Acknowledgments}
The research presented was supported by the Laboratory Directed Research and Development program of Los Alamos National Laboratory under project number 20240004DR. This research was funded in part by the NNSA (DE-NA0003841, DE-NA0003996, DE-NA0004076, DE-NA0004150, DE-FG02-93ER40773, CENTAUR), DOE (NE00009127), NSF (PHY-2310059) and the European Union (ChETEC-INFRA, 101008324). This research was supported by the Department of Energy National Nuclear Security Administration Office of Research, Development, Test, and Evaluation. We are grateful to the staff at the TAMU Cyclotron Institute for providing the high quality beams that made this work possible and the University of Notre Dame for supporting the NSL.





\newpage

\begin{thebibliography}{36}
\ifx \bisbn   \undefined \def \bisbn  #1{ISBN #1}\fi
\ifx \binits  \undefined \def \binits#1{#1}\fi
\ifx \bauthor  \undefined \def \bauthor#1{#1}\fi
\ifx \batitle  \undefined \def \batitle#1{#1}\fi
\ifx \bjtitle  \undefined \def \bjtitle#1{#1}\fi
\ifx \bvolume  \undefined \def \bvolume#1{\textbf{#1}}\fi
\ifx \byear  \undefined \def \byear#1{#1}\fi
\ifx \bissue  \undefined \def \bissue#1{#1}\fi
\ifx \bfpage  \undefined \def \bfpage#1{#1}\fi
\ifx \blpage  \undefined \def \blpage #1{#1}\fi
\ifx \burl  \undefined \def \burl#1{\textsf{#1}}\fi
\ifx \doiurl  \undefined \def \doiurl#1{\url{https://doi.org/#1}}\fi
\ifx \betal  \undefined \def \betal{\textit{et al.}}\fi
\ifx \binstitute  \undefined \def \binstitute#1{#1}\fi
\ifx \binstitutionaled  \undefined \def \binstitutionaled#1{#1}\fi
\ifx \bctitle  \undefined \def \bctitle#1{#1}\fi
\ifx \beditor  \undefined \def \beditor#1{#1}\fi
\ifx \bpublisher  \undefined \def \bpublisher#1{#1}\fi
\ifx \bbtitle  \undefined \def \bbtitle#1{#1}\fi
\ifx \bedition  \undefined \def \bedition#1{#1}\fi
\ifx \bseriesno  \undefined \def \bseriesno#1{#1}\fi
\ifx \blocation  \undefined \def \blocation#1{#1}\fi
\ifx \bsertitle  \undefined \def \bsertitle#1{#1}\fi
\ifx \bsnm \undefined \def \bsnm#1{#1}\fi
\ifx \bsuffix \undefined \def \bsuffix#1{#1}\fi
\ifx \bparticle \undefined \def \bparticle#1{#1}\fi
\ifx \barticle \undefined \def \barticle#1{#1}\fi
\bibcommenthead
\ifx \bconfdate \undefined \def \bconfdate #1{#1}\fi
\ifx \botherref \undefined \def \botherref #1{#1}\fi
\ifx \url \undefined \def \url#1{\textsf{#1}}\fi
\ifx \bchapter \undefined \def \bchapter#1{#1}\fi
\ifx \bbook \undefined \def \bbook#1{#1}\fi
\ifx \bcomment \undefined \def \bcomment#1{#1}\fi
\ifx \oauthor \undefined \def \oauthor#1{#1}\fi
\ifx \citeauthoryear \undefined \def \citeauthoryear#1{#1}\fi
\ifx \endbibitem  \undefined \def \endbibitem {}\fi
\ifx \bconflocation  \undefined \def \bconflocation#1{#1}\fi
\ifx \arxivurl  \undefined \def \arxivurl#1{\textsf{#1}}\fi
\csname PreBibitemsHook\endcsname

\bibitem{AKW99b}
\begin{barticle}
\bauthor{\bsnm{Arlandini}, \binits{C.}},
\bauthor{\bsnm{K{{\"a}}ppeler}, \binits{F.}},
\bauthor{\bsnm{Wisshak}, \binits{K.}},
\bauthor{\bsnm{Gallino}, \binits{R.}},
\bauthor{\bsnm{Lugaro}, \binits{M.}},
\bauthor{\bsnm{Busso}, \binits{M.}},
\bauthor{\bsnm{Straniero}, \binits{O.}}:
\batitle{Neutron capture in low mass asymptotic giant branch stars: cross
  sections and abundance signatures.}
\bjtitle{Ap. J.}
\bvolume{525},
\bfpage{886}--\blpage{900}
(\byear{1999})
\end{barticle}
\endbibitem

\bibitem{WVA01}
\begin{barticle}
\bauthor{\bsnm{Wisshak}, \binits{K.}},
\bauthor{\bsnm{Voss}, \binits{F.}},
\bauthor{\bsnm{Arlandini}, \binits{C.}},
\bauthor{\bsnm{Be\v{c}v\'{a}\v{r}}, \binits{F.}},
\bauthor{\bsnm{Straniero}, \binits{O.}},
\bauthor{\bsnm{Gallino}, \binits{R.}},
\bauthor{\bsnm{Heil}, \binits{M.}},
\bauthor{\bsnm{K{{\"a}}ppeler}, \binits{F.}},
\bauthor{\bsnm{Krti\v{c}ka}, \binits{M.}},
\bauthor{\bsnm{Masera}, \binits{S.}},
\bauthor{\bsnm{Reifarth}, \binits{R.}},
\bauthor{\bsnm{Travaglio}, \binits{C.}}:
\batitle{Neutron capture on $^{180}$ta$^m$: Clue for an $s$-process origin of
  nature's rarest isotope}.
\bjtitle{Physical Review Letters}
\bvolume{87}(\bissue{25}),
\bfpage{251102}
(\byear{2001})
\end{barticle}
\endbibitem

\bibitem{RKV04}
\begin{barticle}
\bauthor{\bsnm{{Reifarth}}, \binits{R.}},
\bauthor{\bsnm{{K{{\"a}}ppeler}}, \binits{F.}},
\bauthor{\bsnm{{Voss}}, \binits{F.}},
\bauthor{\bsnm{{Wisshak}}, \binits{K.}},
\bauthor{\bsnm{{Gallino}}, \binits{R.}},
\bauthor{\bsnm{{Pignatari}}, \binits{M.}},
\bauthor{\bsnm{{Straniero}}, \binits{O.}}:
\batitle{{$^{128}$Xe and $^{130}$Xe: Testing He-Shell Burning in Asymptotic
  Giant Branch Stars}}.
\bjtitle{Ap. J.}
\bvolume{614},
\bfpage{363}
(\byear{2004}).
\doiurl{10.1086/422206}
\end{barticle}
\endbibitem

\bibitem{KGB11}
\begin{barticle}
\bauthor{\bsnm{K\"appeler}, \binits{F.}},
\bauthor{\bsnm{Gallino}, \binits{R.}},
\bauthor{\bsnm{Bisterzo}, \binits{S.}},
\bauthor{\bsnm{Aoki}, \binits{W.}}:
\batitle{The $s$ process: Nuclear physics, stellar models, and observations}.
\bjtitle{Rev. Mod. Phys.}
\bvolume{83},
\bfpage{157}--\blpage{193}
(\byear{2011}).
\doiurl{10.1103/RevModPhys.83.157}
\end{barticle}
\endbibitem

\bibitem{RAH03}
\begin{barticle}
\bauthor{\bsnm{Reifarth}, \binits{R.}},
\bauthor{\bsnm{Arlandini}, \binits{C.}},
\bauthor{\bsnm{Heil}, \binits{M.}},
\bauthor{\bsnm{K{{\"a}}ppeler}, \binits{F.}},
\bauthor{\bsnm{Sedychev}, \binits{P.V.}},
\bauthor{\bsnm{Mengoni}, \binits{A.}},
\bauthor{\bsnm{Herman}, \binits{M.}},
\bauthor{\bsnm{Rauscher}, \binits{T.}},
\bauthor{\bsnm{Gallino}, \binits{R.}},
\bauthor{\bsnm{Travaglio}, \binits{C.}}:
\batitle{Stellar neutron capture on promethium - implications for the
  $s$-process neutron density}.
\bjtitle{Astrophysical Journal}
\bvolume{582},
\bfpage{1251}
(\byear{2003})
\end{barticle}
\endbibitem

\bibitem{KTG16}
\begin{barticle}
\bauthor{\bsnm{Koloczek}, \binits{A.}},
\bauthor{\bsnm{Thomas}, \binits{B.}},
\bauthor{\bsnm{Glorius}, \binits{J.}},
\bauthor{\bsnm{Plag}, \binits{R.}},
\bauthor{\bsnm{Pignatari}, \binits{M.}},
\bauthor{\bsnm{Reifarth}, \binits{R.}},
\bauthor{\bsnm{Ritter}, \binits{C.}},
\bauthor{\bsnm{Schmidt}, \binits{S.}},
\bauthor{\bsnm{Sonnabend}, \binits{K.}}:
\batitle{Sensitivity study for s process nucleosynthesis in agb stars}.
\bjtitle{Atomic Data and Nuclear Data Tables}
\bvolume{108},
\bfpage{1}--\blpage{14}
(\byear{2016}).
\doiurl{10.1016/j.adt.2015.12.001}
\end{barticle}
\endbibitem

\bibitem{sensitivities_online}
\begin{botherref}
\oauthor{\bsnm{Reifarth}, \binits{R.}},
\oauthor{\bsnm{Koloczek}, \binits{A.}},
\oauthor{\bsnm{Plag}, \binits{R.}}:
Sensitivities.
\url{https://exp-astro.de/sensitivities/}
(2020)
\end{botherref}
\endbibitem

\bibitem{HPW11}
\begin{barticle}
\bauthor{\bsnm{{Herwig}}, \binits{F.}},
\bauthor{\bsnm{{Pignatari}}, \binits{M.}},
\bauthor{\bsnm{{Woodward}}, \binits{P.R.}},
\bauthor{\bsnm{{Porter}}, \binits{D.H.}},
\bauthor{\bsnm{{Rockefeller}}, \binits{G.}},
\bauthor{\bsnm{{Fryer}}, \binits{C.L.}},
\bauthor{\bsnm{{Bennett}}, \binits{M.}},
\bauthor{\bsnm{{Hirschi}}, \binits{R.}}:
\batitle{{Convective-reactive Proton-$^{12}$C Combustion in Sakurai's Object
  (V4334 Sagittarii) and Implications for the Evolution and Yields from the
  First Generations of Stars}}.
\bjtitle{Ap. J.}
\bvolume{727},
\bfpage{89}
(\byear{2011})
\end{barticle}
\endbibitem

\bibitem{GZY13}
\begin{barticle}
\bauthor{\bsnm{{Garc{\'{\i}}a-Hern{\'a}ndez}}, \binits{D.A.}},
\bauthor{\bsnm{{Zamora}}, \binits{O.}},
\bauthor{\bsnm{{Yag{\"u}e}}, \binits{A.}},
\bauthor{\bsnm{{Uttenthaler}}, \binits{S.}},
\bauthor{\bsnm{{Karakas}}, \binits{A.I.}},
\bauthor{\bsnm{{Lugaro}}, \binits{M.}},
\bauthor{\bsnm{{Ventura}}, \binits{P.}},
\bauthor{\bsnm{{Lambert}}, \binits{D.L.}}:
\batitle{{Hot bottom burning and s-process nucleosynthesis in massive AGB stars
  at the beginning of the thermally-pulsing phase}}.
\bjtitle{\aap}
\bvolume{555},
\bfpage{3}
(\byear{2013})
\end{barticle}
\endbibitem

\bibitem{SuE01}
\begin{barticle}
\bauthor{\bsnm{Surman}, \binits{R.}},
\bauthor{\bsnm{Engel}, \binits{J.}}:
\batitle{Changes in r-process abundances at late times}.
\bjtitle{Phys. Rev. C}
\bvolume{64},
\bfpage{035801}
(\byear{2001})
\end{barticle}
\endbibitem

\bibitem{RLK14}
\begin{barticle}
\bauthor{\bsnm{{Reifarth}}, \binits{R.}},
\bauthor{\bsnm{{Lederer}}, \binits{C.}},
\bauthor{\bsnm{{K{{\"a}}ppeler}}, \binits{F.}}:
\batitle{{Neutron reactions in astrophysics}}.
\bjtitle{Journal of Physics G Nuclear Physics}
\bvolume{41}(\bissue{5}),
\bfpage{053101}
(\byear{2014})
\end{barticle}
\endbibitem

\bibitem{CoR07}
\begin{barticle}
\bauthor{\bsnm{{Couture}}, \binits{A.}},
\bauthor{\bsnm{{Reifarth}}, \binits{R.}}:
\batitle{{Direct measurements of neutron capture on radioactive isotopes}}.
\bjtitle{Atomic Data and Nuclear Data Tables}
\bvolume{93},
\bfpage{807}
(\byear{2007})
\end{barticle}
\endbibitem

\bibitem{LLK23}
\begin{barticle}
\bauthor{\bsnm{{Laird}}, \binits{A.M.}},
\bauthor{\bsnm{{Lugaro}}, \binits{M.}},
\bauthor{\bsnm{{Kankainen}}, \binits{A.}},
\bauthor{\bsnm{{Adsley}}, \binits{P.}},
\bauthor{\bsnm{{Bardayan}}, \binits{D.W.}},
\bauthor{\bsnm{{Brinkman}}, \binits{H.E.}},
\bauthor{\bsnm{{C{\^o}t{\'e}}}, \binits{B.}},
\bauthor{\bsnm{{Deibel}}, \binits{C.M.}},
\bauthor{\bsnm{{Diehl}}, \binits{R.}},
\bauthor{\bsnm{{Hammache}}, \binits{F.}},
\bauthor{\bsnm{{den Hartogh}}, \binits{J.W.}},
\bauthor{\bsnm{{Jos{\'e}}}, \binits{J.}},
\bauthor{\bsnm{{Kurtulgil}}, \binits{D.}},
\bauthor{\bsnm{{Lederer-Woods}}, \binits{C.}},
\bauthor{\bsnm{{Lotay}}, \binits{G.}},
\bauthor{\bsnm{{Meynet}}, \binits{G.}},
\bauthor{\bsnm{{Palmerini}}, \binits{S.}},
\bauthor{\bsnm{{Pignatari}}, \binits{M.}},
\bauthor{\bsnm{{Reifarth}}, \binits{R.}},
\bauthor{\bsnm{{de S{\'e}r{\'e}ville}}, \binits{N.}},
\bauthor{\bsnm{{Sieverding}}, \binits{A.}},
\bauthor{\bsnm{{Stancliffe}}, \binits{R.J.}},
\bauthor{\bsnm{{Trueman}}, \binits{T.C.L.}},
\bauthor{\bsnm{{Lawson}}, \binits{T.}},
\bauthor{\bsnm{{Vink}}, \binits{J.S.}},
\bauthor{\bsnm{{Massimi}}, \binits{C.}},
\bauthor{\bsnm{{Mengoni}}, \binits{A.}}:
\batitle{{Progress on nuclear reaction rates affecting the stellar production
  of $^{26}$Al}}.
\bjtitle{Journal of Physics G Nuclear Physics}
\bvolume{50}(\bissue{3}),
\bfpage{033002}
(\byear{2023}).
\doiurl{10.1088/1361-6471/ac9cf8}
\end{barticle}
\endbibitem

\bibitem{DGL25}
\begin{barticle}
\bauthor{\bsnm{{Dellmann}}, \binits{S.F.}},
\bauthor{\bsnm{{Glorius}}, \binits{J.}},
\bauthor{\bsnm{{Litvinov}}, \binits{Y.A.}},
\bauthor{\bsnm{{Reifarth}}, \binits{R.}},
\bauthor{\bsnm{{Varga}}, \binits{L.}},
\bauthor{\bsnm{{Aliotta}}, \binits{M.}},
\bauthor{\bsnm{{Amjad}}, \binits{F.}},
\bauthor{\bsnm{{Blaum}}, \binits{K.}},
\bauthor{\bsnm{{Bott}}, \binits{L.}},
\bauthor{\bsnm{{Brandau}}, \binits{C.}},
\bauthor{\bsnm{{Br{\"u}ckner}}, \binits{B.}},
\bauthor{\bsnm{{Bruno}}, \binits{C.G.}},
\bauthor{\bsnm{{Chen}}, \binits{R.-J.}},
\bauthor{\bsnm{{Davinson}}, \binits{T.}},
\bauthor{\bsnm{{Dickel}}, \binits{T.}},
\bauthor{\bsnm{{Dillmann}}, \binits{I.}},
\bauthor{\bsnm{{Dmytriev}}, \binits{D.}},
\bauthor{\bsnm{{Erbacher}}, \binits{P.}},
\bauthor{\bsnm{{Forstner}}, \binits{O.}},
\bauthor{\bsnm{{Freire-Fern{\'a}ndez}}, \binits{D.}},
\bauthor{\bsnm{{Geissel}}, \binits{H.}},
\bauthor{\bsnm{{G{\"o}bel}}, \binits{K.}},
\bauthor{\bsnm{{Griffin}}, \binits{C.J.}},
\bauthor{\bsnm{{Grisenti}}, \binits{R.E.}},
\bauthor{\bsnm{{Gumberidze}}, \binits{A.}},
\bauthor{\bsnm{{Haettner}}, \binits{E.}},
\bauthor{\bsnm{{Hagmann}}, \binits{S.}},
\bauthor{\bsnm{{Heftrich}}, \binits{T.}},
\bauthor{\bsnm{{Heil}}, \binits{M.}},
\bauthor{\bsnm{{He{\ss}}}, \binits{R.}},
\bauthor{\bsnm{{Hillenbrand}}, \binits{P.-M.}},
\bauthor{\bsnm{{Hornung}}, \binits{C.}},
\bauthor{\bsnm{{Joseph}}, \binits{R.}},
\bauthor{\bsnm{{Jurado}}, \binits{B.}},
\bauthor{\bsnm{{Kazanseva}}, \binits{E.}},
\bauthor{\bsnm{{Khasawneh}}, \binits{K.}},
\bauthor{\bsnm{{Kn{\"o}bel}}, \binits{R.}},
\bauthor{\bsnm{{Kostyleva}}, \binits{D.}},
\bauthor{\bsnm{{Kozhuharov}}, \binits{C.}},
\bauthor{\bsnm{{Kulikov}}, \binits{I.}},
\bauthor{\bsnm{{Kuzminchuk}}, \binits{N.}},
\bauthor{\bsnm{{Kurtulgil}}, \binits{D.}},
\bauthor{\bsnm{{Langer}}, \binits{C.}},
\bauthor{\bsnm{{Leckenby}}, \binits{G.}},
\bauthor{\bsnm{{Lederer-Woods}}, \binits{C.}},
\bauthor{\bsnm{{Lestinsky}}, \binits{M.}},
\bauthor{\bsnm{{Litvinov}}, \binits{S.}},
\bauthor{\bsnm{{L{\"o}her}}, \binits{B.}},
\bauthor{\bsnm{{Lorentz}}, \binits{B.}},
\bauthor{\bsnm{{Lorenz}}, \binits{E.}},
\bauthor{\bsnm{{Marsh}}, \binits{J.}},
\bauthor{\bsnm{{Menz}}, \binits{E.}},
\bauthor{\bsnm{{Morgenroth}}, \binits{T.}},
\bauthor{\bsnm{{Mukha}}, \binits{I.}},
\bauthor{\bsnm{{Petridis}}, \binits{N.}},
\bauthor{\bsnm{{Popp}}, \binits{U.}},
\bauthor{\bsnm{{Psaltis}}, \binits{A.}},
\bauthor{\bsnm{{Purushothaman}}, \binits{S.}},
\bauthor{\bsnm{{Rocco}}, \binits{E.}},
\bauthor{\bsnm{{Roy}}, \binits{P.}},
\bauthor{\bsnm{{Sanjari}}, \binits{M.S.}},
\bauthor{\bsnm{{Scheidenberger}}, \binits{C.}},
\bauthor{\bsnm{{Sguazzin}}, \binits{M.}},
\bauthor{\bsnm{{Sidhu}}, \binits{R.S.}},
\bauthor{\bsnm{{Spillmann}}, \binits{U.}},
\bauthor{\bsnm{{Steck}}, \binits{M.}},
\bauthor{\bsnm{{St{\"o}hlker}}, \binits{T.}},
\bauthor{\bsnm{{Surzhykov}}, \binits{A.}},
\bauthor{\bsnm{{Swartz}}, \binits{J.A.}},
\bauthor{\bsnm{{Tanaka}}, \binits{Y.}},
\bauthor{\bsnm{{T{\"o}rnqvist}}, \binits{H.}},
\bauthor{\bsnm{{Vescovi}}, \binits{D.}},
\bauthor{\bsnm{{Volknandt}}, \binits{M.}},
\bauthor{\bsnm{{Weick}}, \binits{H.}},
\bauthor{\bsnm{{Weigand}}, \binits{M.}},
\bauthor{\bsnm{{Woods}}, \binits{P.J.}},
\bauthor{\bsnm{{Yamaguchi}}, \binits{T.}},
\bauthor{\bsnm{{Zhao}}, \binits{J.}}:
\batitle{{First Proton-Induced Cross Sections on a Stored Rare Ion Beam:
  Measurement of $^{118}$Te(p,$\gamma$) for Explosive Nucleosynthesis}}.
\bjtitle{\prl}
\bvolume{134}(\bissue{14}),
\bfpage{142701}
(\byear{2025}).
\doiurl{10.1103/PhysRevLett.134.142701}
\end{barticle}
\endbibitem

\bibitem{ReL14}
\begin{barticle}
\bauthor{\bsnm{Reifarth}, \binits{R.}},
\bauthor{\bsnm{Litvinov}, \binits{Y.A.}}:
\batitle{Measurements of neutron-induced reactions in inverse kinematics}.
\bjtitle{Phys. Rev. ST Accel. Beams}
\bvolume{17},
\bfpage{014701}
(\byear{2014})
\end{barticle}
\endbibitem

\bibitem{RGH17}
\begin{barticle}
\bauthor{\bsnm{Reifarth}, \binits{R.}},
\bauthor{\bsnm{G\"obel}, \binits{K.}},
\bauthor{\bsnm{Heftrich}, \binits{T.}},
\bauthor{\bsnm{Weigand}, \binits{M.}},
\bauthor{\bsnm{Jurado}, \binits{B.}},
\bauthor{\bsnm{K\"appeler}, \binits{F.}},
\bauthor{\bsnm{Litvinov}, \binits{Y.A.}}:
\batitle{Spallation-based neutron target for direct studies of neutron-induced
  reactions in inverse kinematics}.
\bjtitle{Phys. Rev. Accel. Beams}
\bvolume{20},
\bfpage{044701}
(\byear{2017})
\end{barticle}
\endbibitem

\bibitem{RBD18}
\begin{barticle}
\bauthor{\bsnm{Reifarth}, \binits{R.}},
\bauthor{\bsnm{Brown}, \binits{D.}},
\bauthor{\bsnm{Dababneh}, \binits{S.}},
\bauthor{\bsnm{Litvinov}, \binits{Y.A.}},
\bauthor{\bsnm{Mosby}, \binits{S.M.}}:
\batitle{Neutron-induced reactions in nuclear astrophysics}.
\bjtitle{Jordan Journal of Physics}
\bvolume{11},
\bfpage{27}
(\byear{2018})
\end{barticle}
\endbibitem

\bibitem{LBR90}
\begin{barticle}
\bauthor{\bsnm{Lisowski}, \binits{P.W.}},
\bauthor{\bsnm{Bowman}, \binits{C.D.}},
\bauthor{\bsnm{Russell}, \binits{G.J.}},
\bauthor{\bsnm{Wender}, \binits{S.A.}}:
\batitle{The los alamos national laboratory spallation neutron sources}.
\bjtitle{Nucl. Sci. Engineering}
\bvolume{106},
\bfpage{208}
(\byear{1990})
\end{barticle}
\endbibitem

\bibitem{CMR24}
\begin{bchapter}
\bauthor{\bsnm{{Cooper}}, \binits{A.L.}},
\bauthor{\bsnm{{Mosby}}, \binits{S.}},
\bauthor{\bsnm{{Reifarth}}, \binits{R.}},
\bauthor{\bsnm{{Couture}}, \binits{A.}},
\bauthor{\bsnm{{Bennett}}, \binits{E.}},
\bauthor{\bsnm{{Gibson}}, \binits{N.}},
\bauthor{\bsnm{{Gorelov}}, \binits{D.}},
\bauthor{\bsnm{{Keith}}, \binits{C.}},
\bauthor{\bsnm{{Lovell}}, \binits{A.}},
\bauthor{\bsnm{{Misch}}, \binits{G.}},
\bauthor{\bsnm{{Mumpower}}, \binits{M.}}:
\bctitle{{A high-intensity, low-energy heavy ion source for a neutron target
  proof-of-principle experiment at LANSCE}}.
\bsertitle{Journal of Physics Conference Series},
vol. \bseriesno{2743},
p. \bfpage{012091}
(\byear{2024}).
\doiurl{10.1088/1742-6596/2743/1/012091}
\end{bchapter}
\endbibitem

\bibitem{ISNAP}
\begin{botherref}
Institute for Structure and Nuclear Astrophysics (ISNAP) at the University of
  Notre Dame.
\url{https://isnap.nd.edu/}
(2024)
\end{botherref}
\endbibitem

\bibitem{CI-TAMU}
\begin{botherref}
Cyclotron Institute, Texas A\&M University.
\url{https://cyclotron.tamu.edu/}
(2024)
\end{botherref}
\endbibitem

\bibitem{GEA93}
\begin{botherref}
\oauthor{\bsnm{Apostolakis}, \binits{J.}}:
Cern program library long writeup, w5013.
Technical report,
CERN, GEANT library
(1993).
http://wwwinfo.cern.ch/asd/geant/
\end{botherref}
\endbibitem

\bibitem{CRC24}
\begin{botherref}
\oauthor{\bsnm{Cantrell}, \binits{O.R.}},
\oauthor{\bsnm{Reifarth}, \binits{R.}},
\oauthor{\bsnm{Couture}, \binits{A.J.}},
\oauthor{\bsnm{Cooper}, \binits{A.L.}}:
Carbon block density.
Technical report,
LA-UR-24-27599, Los Alamos National Laboratory (LANL), Los Alamos, NM (United
  States)
(July 2024).
\doiurl{10.2172/2406679}.
\url{https://www.osti.gov/biblio/2406679}
\end{botherref}
\endbibitem

\bibitem{AWB22}
\begin{barticle}
\bauthor{\bsnm{Arregui-Mena}, \binits{J.D.}},
\bauthor{\bsnm{Worth}, \binits{R.N.}},
\bauthor{\bsnm{Bodel}, \binits{W.}},
\bauthor{\bsnm{M\"arz}, \binits{B.}},
\bauthor{\bsnm{Li}, \binits{W.}},
\bauthor{\bsnm{Campbell}, \binits{A.A.}},
\bauthor{\bsnm{Cakmak}, \binits{E.}},
\bauthor{\bsnm{Gallego}, \binits{N.}},
\bauthor{\bsnm{Contescu}, \binits{C.}},
\bauthor{\bsnm{Edmondson}, \binits{P.D.}}:
\batitle{Multiscale characterization and comparison of historical and modern
  nuclear graphite grades}.
\bjtitle{Materials Characterization}
\bvolume{190},
\bfpage{112047}
(\byear{2022}).
\doiurl{10.1016/j.matchar.2022.112047}
\end{barticle}
\endbibitem

\bibitem{ChS21}
\begin{botherref}
\oauthor{\bsnm{Chen}, \binits{J.}},
\oauthor{\bsnm{Singh}, \binits{B.}}:
Nuclear data sheets for a=194.
Nuclear Data Sheets
\textbf{177}
(2021).
\doiurl{10.1016/j.nds.2021.09.001}
\end{botherref}
\endbibitem

\bibitem{Xia07}
\begin{botherref}
\oauthor{\bsnm{Xiaolong}, \binits{H.}}:
Nuclear data sheets for a = 196.
Nuclear Data Sheets
\textbf{108}
(2007).
\doiurl{10.1016/j.nds.2007.05.001}
\end{botherref}
\endbibitem

\bibitem{XiM16}
\begin{barticle}
\bauthor{\bsnm{Xiaolong}, \binits{H.}},
\bauthor{\bsnm{Mengxiao}, \binits{K.}}:
\batitle{Nuclear data sheets for a = 198}.
\bjtitle{Nuclear Data Sheets}
\bvolume{133},
\bfpage{221}--\blpage{416}
(\byear{2016}).
\doiurl{10.1016/j.nds.2016.02.002}
\end{barticle}
\endbibitem

\bibitem{TCG02}
\begin{barticle}
\bauthor{\bsnm{{Tilley}}, \binits{D.R.}},
\bauthor{\bsnm{{Cheves}}, \binits{C.M.}},
\bauthor{\bsnm{{Godwin}}, \binits{J.L.}},
\bauthor{\bsnm{{Hale}}, \binits{G.M.}},
\bauthor{\bsnm{{Hofmann}}, \binits{H.M.}},
\bauthor{\bsnm{{Kelley}}, \binits{J.H.}},
\bauthor{\bsnm{{Sheu}}, \binits{C.G.}},
\bauthor{\bsnm{{Weller}}, \binits{H.R.}}:
\batitle{{Energy levels of light nuclei /A=5, 6, 7}}.
\bjtitle{\nphysa}
\bvolume{708}(\bissue{1}),
\bfpage{3}--\blpage{163}
(\byear{2002}).
\doiurl{10.1016/S0375-9474(02)00597-3}
\end{barticle}
\endbibitem

\bibitem{REF18}
\begin{barticle}
\bauthor{\bsnm{{Reifarth, Ren\'e}}},
\bauthor{\bsnm{{Erbacher, Philipp}}},
\bauthor{\bsnm{{Fiebiger, Stefan}}},
\bauthor{\bsnm{{G\"obel, Kathrin}}},
\bauthor{\bsnm{{Heftrich, Tanja}}},
\bauthor{\bsnm{{Heil, Michael}}},
\bauthor{\bsnm{{K\"appeler, Franz}}},
\bauthor{\bsnm{{Klapper, Nadine}}},
\bauthor{\bsnm{{Kurtulgil, Deniz}}},
\bauthor{\bsnm{{Langer, Christoph}}},
\bauthor{\bsnm{{Lederer-Woods, Claudia}}},
\bauthor{\bsnm{{Mengoni, Alberto}}},
\bauthor{\bsnm{{Thomas, Benedikt}}},
\bauthor{\bsnm{{Schmidt, Stefan}}},
\bauthor{\bsnm{{Weigand, Mario}}},
\bauthor{\bsnm{{Wiescher, Michael}}}:
\batitle{Neutron-induced cross sections - from raw data to astrophysical
  rates}.
\bjtitle{Eur. Phys. J. Plus}
\bvolume{133}(\bissue{10}),
\bfpage{424}
(\byear{2018}).
\doiurl{10.1140/epjp/i2018-12295-3}
\end{barticle}
\endbibitem

\bibitem{RHK09}
\begin{barticle}
\bauthor{\bsnm{{Reifarth}}, \binits{R.}},
\bauthor{\bsnm{{Heil}}, \binits{M.}},
\bauthor{\bsnm{{K{{\"a}}ppeler}}, \binits{F.}},
\bauthor{\bsnm{{Plag}}, \binits{R.}}:
\batitle{{PINO-a tool for simulating neutron spectra resulting from the
  $^7$Li(p,n) reaction}}.
\bjtitle{Nucl. Inst. Meth. A}
\bvolume{608},
\bfpage{139}
(\byear{2009})
\end{barticle}
\endbibitem

\bibitem{pino_online}
\begin{botherref}
\oauthor{\bsnm{Reifarth}, \binits{R.}},
\oauthor{\bsnm{Plag}, \binits{R.}},
\oauthor{\bsnm{Erbacher}, \binits{P.}}:
PINO - neutron spectra from $^{7}$Li(p,n) and $^{9}$Be(p,n).
\url{https://exp-astro.de/pino/}
(2020)
\end{botherref}
\endbibitem

\bibitem{BCC18}
\begin{barticle}
\bauthor{\bsnm{{Brown}}, \binits{D.A.}},
\bauthor{\bsnm{{Chadwick}}, \binits{M.B.}},
\bauthor{\bsnm{{Capote}}, \binits{R.}},
\bauthor{\bsnm{{Kahler}}, \binits{A.C.}},
\bauthor{\bsnm{{Trkov}}, \binits{A.}},
\bauthor{\bsnm{{Herman}}, \binits{M.W.}},
\bauthor{\bsnm{{Sonzogni}}, \binits{A.A.}},
\bauthor{\bsnm{{Danon}}, \binits{Y.}},
\bauthor{\bsnm{{Carlson}}, \binits{A.D.}},
\bauthor{\bsnm{{Dunn}}, \binits{M.}},
\bauthor{\bsnm{{Smith}}, \binits{D.L.}},
\bauthor{\bsnm{{Hale}}, \binits{G.M.}},
\bauthor{\bsnm{{Arbanas}}, \binits{G.}},
\bauthor{\bsnm{{Arcilla}}, \binits{R.}},
\bauthor{\bsnm{{Bates}}, \binits{C.R.}},
\bauthor{\bsnm{{Beck}}, \binits{B.}},
\bauthor{\bsnm{{Becker}}, \binits{B.}},
\bauthor{\bsnm{{Brown}}, \binits{F.}},
\bauthor{\bsnm{{Casperson}}, \binits{R.J.}},
\bauthor{\bsnm{{Conlin}}, \binits{J.}},
\bauthor{\bsnm{{Cullen}}, \binits{D.E.}},
\bauthor{\bsnm{{Descalle}}, \binits{M.-A.}},
\bauthor{\bsnm{{Firestone}}, \binits{R.}},
\bauthor{\bsnm{{Gaines}}, \binits{T.}},
\bauthor{\bsnm{{Guber}}, \binits{K.H.}},
\bauthor{\bsnm{{Hawari}}, \binits{A.I.}},
\bauthor{\bsnm{{Holmes}}, \binits{J.}},
\bauthor{\bsnm{{Johnson}}, \binits{T.D.}},
\bauthor{\bsnm{{Kawano}}, \binits{T.}},
\bauthor{\bsnm{{Kiedrowski}}, \binits{B.C.}},
\bauthor{\bsnm{{Koning}}, \binits{A.J.}},
\bauthor{\bsnm{{Kopecky}}, \binits{S.}},
\bauthor{\bsnm{{Leal}}, \binits{L.}},
\bauthor{\bsnm{{Lestone}}, \binits{J.P.}},
\bauthor{\bsnm{{Lubitz}}, \binits{C.}},
\bauthor{\bsnm{{M{\'a}rquez Dami{\'a}n}}, \binits{J.I.}},
\bauthor{\bsnm{{Mattoon}}, \binits{C.M.}},
\bauthor{\bsnm{{McCutchan}}, \binits{E.A.}},
\bauthor{\bsnm{{Mughabghab}}, \binits{S.}},
\bauthor{\bsnm{{Navratil}}, \binits{P.}},
\bauthor{\bsnm{{Neudecker}}, \binits{D.}},
\bauthor{\bsnm{{Nobre}}, \binits{G.P.A.}},
\bauthor{\bsnm{{Noguere}}, \binits{G.}},
\bauthor{\bsnm{{Paris}}, \binits{M.}},
\bauthor{\bsnm{{Pigni}}, \binits{M.T.}},
\bauthor{\bsnm{{Plompen}}, \binits{A.J.}},
\bauthor{\bsnm{{Pritychenko}}, \binits{B.}},
\bauthor{\bsnm{{Pronyaev}}, \binits{V.G.}},
\bauthor{\bsnm{{Roubtsov}}, \binits{D.}},
\bauthor{\bsnm{{Rochman}}, \binits{D.}},
\bauthor{\bsnm{{Romano}}, \binits{P.}},
\bauthor{\bsnm{{Schillebeeckx}}, \binits{P.}},
\bauthor{\bsnm{{Simakov}}, \binits{S.}},
\bauthor{\bsnm{{Sin}}, \binits{M.}},
\bauthor{\bsnm{{Sirakov}}, \binits{I.}},
\bauthor{\bsnm{{Sleaford}}, \binits{B.}},
\bauthor{\bsnm{{Sobes}}, \binits{V.}},
\bauthor{\bsnm{{Soukhovitskii}}, \binits{E.S.}},
\bauthor{\bsnm{{Stetcu}}, \binits{I.}},
\bauthor{\bsnm{{Talou}}, \binits{P.}},
\bauthor{\bsnm{{Thompson}}, \binits{I.}},
\bauthor{\bsnm{{van der Marck}}, \binits{S.}},
\bauthor{\bsnm{{Welser-Sherrill}}, \binits{L.}},
\bauthor{\bsnm{{Wiarda}}, \binits{D.}},
\bauthor{\bsnm{{White}}, \binits{M.}},
\bauthor{\bsnm{{Wormald}}, \binits{J.L.}},
\bauthor{\bsnm{{Wright}}, \binits{R.Q.}},
\bauthor{\bsnm{{Zerkle}}, \binits{M.}},
\bauthor{\bsnm{{{\v Z}erovnik}}, \binits{G.}},
\bauthor{\bsnm{{Zhu}}, \binits{Y.}}:
\batitle{{ENDF/B-VIII.0: The 8$^{th}$ Major Release of the Nuclear Reaction
  Data Library with CIELO-project Cross Sections, New Standards and Thermal
  Scattering Data}}.
\bjtitle{Nuclear Data Sheets}
\bvolume{148},
\bfpage{1}--\blpage{142}
(\byear{2018})
\end{barticle}
\endbibitem

\bibitem{RCO25}
\begin{botherref}
\oauthor{\bsnm{Reifarth}, \binits{R.}},
\oauthor{\bsnm{Couture}, \binits{A.J.}},
\oauthor{\bsnm{O'Brien}, \binits{E.M.}},
\oauthor{\bsnm{Vermeulen}, \binits{C.}},
\oauthor{\bsnm{Cooper}, \binits{A.L.}},
\oauthor{\bsnm{Dellmann}, \binits{S.F.}},
\oauthor{\bsnm{Knapova}, \binits{I.}},
\oauthor{\bsnm{Mosby}, \binits{S.M.}}:
Neutron moderation at ipf.
Technical report,
LA-UR-25-23900, Los Alamos National Laboratory (LANL), Los Alamos, NM (United
  States)
(April 2025).
\doiurl{10.2172/2561261}.
\url{https://www.osti.gov/biblio/2561261}
\end{botherref}
\endbibitem

\bibitem{ZMK18}
\begin{barticle}
\bauthor{\bsnm{{Zavorka}}, \binits{L.}},
\bauthor{\bsnm{{Mocko}}, \binits{M.J.}},
\bauthor{\bsnm{{Koehler}}, \binits{P.E.}}:
\batitle{{Physics design of the next-generation spallation neutron
  target-moderator-reflector-shield assembly at LANSCE}}.
\bjtitle{Nuclear Instruments and Methods in Physics Research A}
\bvolume{901},
\bfpage{189}--\blpage{197}
(\byear{2018}).
\doiurl{10.1016/j.nima.2018.06.018}
\end{barticle}
\endbibitem

\bibitem{RCC23}
\begin{botherref}
\oauthor{\bsnm{Reifarth}, \binits{R.}},
\oauthor{\bsnm{Cooper}, \binits{A.L.}},
\oauthor{\bsnm{Couture}, \binits{A.}},
\oauthor{\bsnm{Mosby}, \binits{S.M.}}:
Cubical moderators for the neutron target demonstrator (ntd).
Technical report,
Report LA-UR-23-33899, Los Alamos National Laboratory
(2023).
\doiurl{10.2172/2229692}.
\url{https://www.osti.gov/biblio/2229692}
\end{botherref}
\endbibitem

\bibitem{RHO05}
\begin{barticle}
\bauthor{\bsnm{{Rochman}}, \binits{D.}},
\bauthor{\bsnm{{Haight}}, \binits{R.C.}},
\bauthor{\bsnm{{O' Donnell}}, \binits{J.M.}},
\bauthor{\bsnm{{Michaudon}}, \binits{A.}},
\bauthor{\bsnm{{Wender}}, \binits{S.A.}},
\bauthor{\bsnm{{Vieira}}, \binits{D.J.}},
\bauthor{\bsnm{{Bond}}, \binits{E.M.}},
\bauthor{\bsnm{{Bredeweg}}, \binits{T.A.}},
\bauthor{\bsnm{{Kronenberg}}, \binits{A.}},
\bauthor{\bsnm{{Wilhelmy}}, \binits{J.B.}},
\bauthor{\bsnm{{Ethvignot}}, \binits{T.}},
\bauthor{\bsnm{{Granier}}, \binits{T.}},
\bauthor{\bsnm{{Petit}}, \binits{M.}},
\bauthor{\bsnm{{Danon}}, \binits{Y.}}:
\batitle{{Characteristics of a lead slowing-down spectrometer coupled to the
  LANSCE accelerator}}.
\bjtitle{Nuclear Instruments and Methods in Physics Research A}
\bvolume{550}(\bissue{1-2}),
\bfpage{397}--\blpage{413}
(\byear{2005}).
\doiurl{10.1016/j.nima.2005.04.075}
\end{barticle}
\endbibitem

\end{thebibliography}


\end{document}